\documentclass[14pt,prb ,article]{revtex4-1}

\usepackage{amsmath}  
\usepackage{amsfonts} 
\usepackage{graphicx} 
\usepackage{enumitem} 
\setitemize{noitemsep,topsep=0pt,parsep=0pt,partopsep=0pt}
\usepackage{bm}    
\usepackage{hyperref}    
\usepackage{extsizes}    
\usepackage{setspace}
\usepackage[margin=70pt]{geometry}

\begin{document}

\title{The Bones of Sophus Lie}

\author{Clinton L. Lewis}
\affiliation{Division of Science and Mathematics, West Valley College, Saratoga, California (retired)}
\email{CLewis@Swiftwords.com}

\date{\today}

\begin{abstract}
The gauge covariant derivative of a wave function is ubiquitous in gauge theory, and with associated gauge transformations it defines charged currents interacting with external fields, such as the Lorentz force exerted by an electromagnetic field. It is the gauge covariant derivative which defines how an external field acts upon the wave function. This paper constructs the gauge covariant derivative, then uses the elegant framework of Lagrangian mechanics to derive two ``divergence equations'' from a general Lagrangian, one applying to the charged current, the other to energy-momentum. The student will appreciate the construction of the gauge covariant derivative of a classical wave function using only matrices, linear transformations, external fields, and partial derivatives. More unusual is using the principle of covariance, rather than group theory as guidance in the construction, but with exactly the same result. Advantage is taken of the close analogy with coordinate covariance of tensors. The details of deriving these two divergences provides motivation and a path to understanding the gauge covariant derivative, the underlying non-abelian Lie algebra, its application to building Lagrangians, and the resulting definitions of charged current and energy-momentum. All results are generalized to curved space-time.
\end{abstract}

\maketitle

\tableofcontents

\newpage

\section{Introduction}

We study how a classical wave function, its gauge covariant derivative, and the corresponding Lagrangian respond to a gauge transformation, then conspire to define a charged current, the divergence of that current, and an external Lorentz force acting upon the energy-momentum of the system. These results appear as two ``divergence-type'' equations, one for charge, the other for energy-momentum.

These are ancient topics in gauge theory\cite{AitchisonP28}, and Lagrangian mechanics, but the presentation here is shorted by using the principle of covariance as the central guiding principle. Covariance is the response of the wave function to a linear homogeneous transformation. The object of this paper is to construct from first principles an essential tool in building Lagrangians, the \textit{gauge covariant derivative} of a classical wave function in curved space-time. In the simplest case only two objects enter the Lagrangian, the wave function and its derivative.

The hope is to appeal to students with a presentation of the mathematical properties of the gauge covariant derivative, but closely motivated by physics in the form of the two divergences, and using only elements of differential equations, matrix linear algebra, and tensor index notation. 

The gauge covariant derivative appears in Lagrangians which are invariant under the application of a gauge transformation. Such a Lagrangian is said to have a  \textit{symmetry} with respect to this transformation. Further, this symmetry implies a charged current with a vanishing divergence. In the end, since all parts defined here have specific gauge transformation properties, the result is a “toolbox” for building gauge invariant Lagrangians.

Consider a  wave function represented as a column matrix ${\bm\phi }$ with complex-valued elements, each a function of position in space-time ${x^\mu }$, $\mu  = 0,1,2,3$.
\begin{equation}
\label{matrixWavefunction}
{\bm\phi }\left( x \right){ = }\left( {\begin{array}{*{20}{c}}
  {{\phi _{\left( 1 \right)}}\left( x \right)} \\ 
   \vdots  \\ 
  {{\phi _{\left( n \right)}}\left( x \right)} 
\end{array}} \right)
\end{equation}
The partial derivative is defined according to the usual limiting process,
\begin{equation}
\label{derivativeLimit}
\frac{{\partial {\bm\phi }\left( x \right)}}{{\partial {x^\mu }}}{ = }\mathop {\lim }\limits_{\Delta {x^\mu } \to 0} \frac{{{\bm\phi }\left( {x + \Delta {x^\mu }} \right) - {\bm\phi }\left( x \right)}}{{\Delta {x^\mu }}}
\end{equation}
The partial derivative applies to each component of the wave function.
\begin{equation}
\label{derivativeMatrix}
\frac{{\partial {\bm\phi }\left( x \right)}}{{\partial {x^\mu }}}{ = }\left( {\begin{array}{*{20}{c}}
  {\frac{{\partial {\phi _1}\left( x \right)}}{{\partial {x^\mu }}}} \\ 
   \vdots  \\ 
  {\frac{{\partial {\phi _n}\left( x \right)}}{{\partial {x^\mu }}}} 
\end{array}} \right)
\end{equation}

A gauge covariant derivative of a wave function is related to the partial derivative, but adds an additional conceptual layer -- that of the linear and homogeneous response of the wave function to a \textit{gauge transformation}. By construction, the gauge covariant derivative of the wave function responds to the \textit{same} gauge transformation.

Implement the gauge transformation by left-multiplying the wave function by the transformation represented as a square matrix ${\mathbf{T}}\left( x \right)$,
\begin{equation}
\label{TMatrix}
{\mathbf{T}}{\bm\phi }
\end{equation}
By construction to be explored, the gauge covariant derivative of the wave function ${D_\mu }\left( {\bm\phi } \right)$   \textit{covaries} which means that the derivative of the wave function responds to the identical transformation as the wave function itself.
\begin{equation}
\label{TDMatrix}
{\mathbf{T}}{D_\mu }\left( {\bm\phi } \right)
\end{equation}
The gauge transformation matrix, to be defined later, is essential to the definition of the gauge covariant derivative.

How to motivate and add linear gauge transformation properties to the ordinary partial derivative to construct a gauge covariant derivative, occupies the majority of this paper.

Analogously\cite{WeinP6}, the \textit{coordinate} covariant derivative used in tensor algebra has linear homogeneous transformation properties under a general coordinate transformation. Historically, general relativity with essential use of coordinate covariant derivatives was introduced in 1915, while gauge transformations were introduced later by Hermann Weyl, and Kibble\cite{Kibble1961} , and still later as part of the Standard Model.

Two applications of the gauge covariant derivative are explored here,
\begin{itemize}
	\item the definition and divergence of a charged current \textit{vector}, and
	\item the definition and divergence of the energy-momentum \textit{tensor} while subject to external Lorentz forces.
\end{itemize}
These ``two divergences'' provide physical motivation for the mathematical properties and construction of the gauge covariant derivative.

The \textbf{stage} for the evolution of fields in our system is a general space-time metric. The metric, although arbitrary, is a fixed background metric so that Einstein equations do not apply. Two fields exist in this space-time, a wave function representing matter, and a generalization of the electromagnetic field acting as an  \textit{external} force.

The \textbf{notation} emphasizes covariance, so tensor notation, Einstein summation and covariant derivatives are used throughout. The development, although covariant, is not general relativistic since the system considered here is subject to a force due to an external gauge field, a situation impossible in general relativity where there are no ``external''  forces.

\textbf{Matrix notation} is used to avoid a paroxysm of indices. \textbf{Bold type} indicates a matrix with accompanying gauge transformation properties.

Derivatives with respect to a \textbf{matrix} are to be evaluated component by component, with matrix multiplication indicating summation of matrix components. A matrix equation may be converted to a component equation by restoring matrix indices. A partial derivative of an invariant scalar, such as the Lagrangian, with respect to a column matrix yields a row matrix.

\section{Continuity equations and conservation}

Continuity equations have intuitive appeal since continuity conveys the feeling of an enduring, ponderable substance, neither created nor destroyed in an isolated system. These continuity equations appear as a vanishing divergence of a vector, hence, our pursuit of the two divergences mentioned above. 

The continuity equation for a fluid of density $P$ and current ${\mathbf{J}}$ at a point in space $\left( {r,t} \right)$ is,\cite{}
\begin{equation}
\label{3DcurrentCons}
  \frac{\partial }{{\partial t}}P\left( {r,t} \right) + \nabla  \cdot {\mathbf{J}}\left( {r,t} \right) = 0
\end{equation}
The continuity equation implies a conservation law in the sense that the quantity ${\mathbf{Q}}$ of fluid contained in a given volume $V$ with an exterior surface ${\mathbf{S}}$ varies only with the passage of current ${\mathbf{J}}$ through the exterior surface.
\begin{equation}
\frac{\partial }{{\partial t}}{\mathbf{Q}} = \frac{\partial }{{\partial t}}\int\limits_V {P\left( {r,t} \right)} dV =  - \int\limits_S {{\mathbf{J}}\left( {r,t} \right)}  \cdot d{\mathbf{S}}
\end{equation}
If the surface is extended, perhaps to infinity, to include all external currents, then the quantity in the volume is constant, so that the derivative with respect to time would be zero. Fluid \textit{conservation} means that no fluid is created or lost within the volume, without having entered or exited through the enclosing surface.
The vanishing ``divergence-type'' expression in Eq.~(\ref{3DcurrentCons}) becomes the following in Minkowski space-time with a diagonal constant metric.
\begin{equation}
  {\partial _\mu }{j^\mu } = 0
\end{equation}
The density $P$ is the time component of the 4 dimensional current vector ${j^\mu }$. In covariant tensor notation, the vanishing divergence  represents a continuity equation for a current in a general curved space-time.\cite{LandauP72,SchiffP468} 
\begin{equation}
  {\nabla _\mu }{j^\mu } = 0
\end{equation}
The vanishing covariant derivative is brought into a form exposing ordinary partial derivatives by using an identity\cite{CarmeliDerivIdent}.   The absolute value of the determinant of the metric ${g^{\mu \nu }}\left( x \right)$ is ``$g$'' in the following tensor expression. This form makes clear the continuity equation consisting of the vanishing divergence of the ordinary partial derivative.
\begin{equation}
\label{divZeroPartial}
  {\nabla _\mu }{j^\mu } = \frac{1}{{\sqrt g }}{\partial _\mu }\left( {\sqrt g {j^\mu }} \right) = 0
\end{equation}
This form of the vanishing covariant divergence is a  generalization of Eq.~(\ref{derivativeMatrix}), and is again an equation of continuity because of the appearance of the ordinary partial derivative. Hence the importance of discussing vanishing divergence of current and energy momentum, and the connection to conservation.

The pursuit of this type of ``local'' conservation law at a point, or continuity equation is one of the motivations for our study of gauge covariant derivatives appearing in a vanishing divergence.

\section{Two divergences, the abelian case}

As an example of the two divergences mentioned above, we explore the wave function in an electromagnetic field.\cite{MessKG} These divergence equations are later generalized by using a more general gauge covariant derivative. In the following, $\hbar  = 1$.

The prototypical gauge transformation is that of electromagnetics and the corresponding phase transformation of the wave function.\cite{PeskinGuageCov}~ \cite{FramptonP7}  This example provides guidance for the more complex non-abelian gauge transformations explored here.

The gauge covariant derivative of the single-component, complex-valued wave function ${D_\mu }\phi \left( x \right)$, a function of space-time position ${x^\mu }$, is the familiar,
\begin{equation}
\label{electroGCD}
  {D_\mu }{\phi } = {\partial _\mu }{\phi } + ie{A_\mu }{\phi }
\end{equation}
where  ${A_\mu }$ is the electromagnetic 4-vector potential, and where ``$e$'' is the ordinary electric charge coupling constant. This defines a \textit{gauge} covariant derivative, because the derivative has the same phase (gauge) transformation properties as the wave function field itself. The derivative ``covaries'' with respect to a phase change indicated by the real parameter $\theta \left( x \right)$ as follows,
\begin{equation}
\label{electroPhase}
  \begin{gathered}
  {\phi }\xrightarrow[G]{}{e^{ie\theta \left( x \right)}}{\phi } \hfill \\
  {D_\mu }{\phi }\xrightarrow[G]{}{e^{ie\theta \left( x \right)}}{D_\mu }{\phi } \hfill \\ 
\end{gathered} 
\end{equation}
where the electromagnetic vector potential must simultaneously gauge transform in correspondence with the phase transformation of the wave function.
\begin{equation}
\label{electroGauge}
  {A_\mu }\xrightarrow[G]{}{A_\mu } - {\partial _\mu }\theta 
\end{equation}
The right arrow ``$\xrightarrow[G]{}$'' indicates ``replacement'' since these are finite transformations. Presented in derivative form, the infinitesimal version of the same gauge transformation is
\begin{equation}
  \begin{gathered}
  \delta {\phi /}\delta \theta { = }ie{\phi } \hfill \\
  \delta {D_\mu }{\phi /}\delta \theta  = ie{D_\mu }{\phi } \hfill \\
  \delta {A_\mu } =  - {\partial _\mu }\left( {\delta \theta } \right) \hfill \\ 
\end{gathered} 
\end{equation}
where $\delta$ is the transformation due to a small change in the parameter $\delta \theta$.

The gauge covariant derivative ${D_\mu }$ does not commute in general.  Define the antisymmetric electromagnetic field tensor ${F_{\mu \nu }}$.
\begin{equation}
  {F_{\mu \nu }} = {\partial _\mu }{A_\nu } - {\partial _\nu }{A_\mu }
\end{equation}
which is gauge transformation \textit{invariant},
\begin{equation}
  {F_{\mu \nu }}\xrightarrow[G]{}{F_{\mu \nu }}
\end{equation}
Computation shows that the commutator of the gauge covariant derivative is\cite{MessCommutation}
\begin{equation}
\label{emComm}
  \left[ {{D_\mu },{D_\nu }} \right]{\phi } = ie{F_{\mu \nu }}{\phi }
\end{equation}

Assume the Lagrangian for the wave function is,\cite{CarrollP386}
\begin{equation}
  L\left( {\phi ,{D_\mu }\phi {,}{g_{\mu \nu }}} \right) = {D^\lambda }{\phi ^*}{D_\lambda }\phi  - {M^2}{\phi ^*}\phi 
\end{equation}
The Lagrangian implies the Euler-Lagrange equation which results in the Klein-Gordon equation.\cite{SakuraiP491}
\begin{equation}
\label{KGequ}
  {D^\lambda }{D_\lambda }\phi  + {M^2}\phi  = 0
\end{equation}
The gauge transformation applied to the invariant Lagrangian defines the current,
\begin{equation}
  {j^\nu } = ie\left( {{\phi ^\dag }{D^\nu }\phi  - \left( {{D^\nu }{\phi ^\dag }} \right)\phi } \right)
\end{equation}
The canonical energy-momentum tensor of the Klein-Gordon scalar field is,
\begin{equation}
  T_{S\mu }^{\;\;\;\nu } = {D_\mu }{{\phi }^\dag }{D^\nu }{\phi } + {D_\mu }{\phi }{D^\nu }{{\phi }^\dag } - \delta _\mu ^\nu L
\end{equation}
These are the subject of the ``two divergences'' referred to earlier.

The divergence charged electric current  ${j^\mu }$ vanishes, and so satisfies the continuity equation.\cite{LandauP77}~ \cite{ShankarP165}.  
\begin{equation}
  \label{divZeroVector}
 {\nabla _\mu }{j^\mu } = 0
\end{equation}

The divergence of the energy-momentum tensor ${T_\mu }^\nu $ for matter equals the Lorentz force,\cite{LandauR21}~ \cite{CarmeliLorentzForce}~ \cite{BarutP139} hence does not vanish in general.
\begin{equation}
\label{emLorentzForce}
  {\nabla _\nu }\left( {{T_\mu }^\nu } \right) = {F_{\mu \nu }}{J^\nu }
\end{equation}
The familiar vector component notation for the electromagnetic field tensor ${F_{\mu \nu }}$ is
\begin{equation}
\label{EMFieldTensor}
{F_{\mu \nu }} = \left( {\begin{array}{*{20}{c}}   0&{ - {E_x}}&{ - {E_y}}&{ - {E_z}} \\    {{E_x}}&0&{{H_z}}&{ - {H_y}} \\    {{E_y}}&{ - {H_z}}&0&{{H_x}} \\    {{E_z}}&{{H_y}}&{ - {H_x}}&0  \end{array}} \right)
\end{equation}

We pursue the non-abelian generalization of the two divergences, Eqs.~ (\ref{divZeroVector}) and (\ref{divZeroPartial}) as an application, as well as motivation for the construction of a more general gauge covariant derivative than Eq.~(\ref{electroGCD}). The exploration provided here will not be specific as to the matrix representation, but instead try to expose larger patterns applying to all representations.

\section{The Lagrangian for a wave function}

Begin with the Lagrangian of a system evolving in space-time with matter represented as a wave function, as in Eq.~(\ref{matrixWavefunction}). The space-time has a general metric ${g^{\mu \nu }}\left( x \right)$. Assume a scalar Lagrangian   dependent upon the space-time coordinates only through the wave function and fixed background metric, so that there is no coordinate ``$x$'' appearing explicitly in the Lagrangian.\cite{Doughty}
\begin{equation}
\label{Lagrangian}
L = L\left( {{\bm\phi },{D_\mu }{\bm\phi },{g^{\mu \nu }}} \right)
\end{equation}
The symbol ``${D_\mu }$'' indicates a \textit{gauge covariant} partial derivative with respect to coordinate $x^\mu$ which, like all derivatives, satisfies the product rule (Leibniz rule). For now simply note that it does not commute as does the ordinary partial derivative with respect to coordinates ${\partial }/{{\partial {x^\mu }}}$, or in short form  ${\partial _\mu }$.

The gauge covariant derivative of the metric is defined to vanish, the \textit{metricity} condition.\cite{BlagoP57} By \textit{definition} the gauge covariant derivative $D_\mu$ becomes to the coordinate covariant derivative $\nabla _\lambda $ upon application to gauge invariant objects such as the metric.
\begin{equation}
\label{metricity} {D_\lambda }{g^{\mu \nu }}\left( x \right) = {\nabla _\lambda }{g^{\mu \nu }}\left( x \right) = 0
\end{equation}

As shown later, the Lagrangian defines equations of motion for the wave function through the application of the Euler-Lagrange equations which then control the evolution of the wave equation.

\section{The gauge transformation of the wave function}

Consider the following implementation of a \textit{gauge transformation} as a homogeneous nonsingular transformation applied to a column matrix ${\bm\phi }$ wave function. Left-multiply the wave function with the complex-valued square matrix ${\mathbf{T}}\left( x \right)$  which is then transformed to another column matrix,  ${\mathbf{T}}{\bm\phi }$, \textit{also} a wave function. In what sense the transformed wave function is the \text{same} type of wave function reaches deep into the meaning of this transformation. It will become clear that the transformation matrix must be drawn from the members of a specific Lie group.

Represent a gauge transformation with the arrow ``$\xrightarrow[g]{}$'' which means ``replace each occurrence'' of the symbol on the left with the expression on the right. Another useful notation uses a ``$\widehat{hat}$'' to indicate the gauge transformed expression. With this notation, the gauge covariant transformation of the wave function is,\cite{RamondLocalTransf}
\begin{equation}
\label{gaugeTransPhi}
{\bm\phi }\xrightarrow[g]{}{{\hat{\bm\phi}}  = {\mathbf{T}}{\bm\phi }}
\end{equation}

A transformation may do nothing, in which case it is the \textit{identity} transformation.

Some objects are gauge invariant, such as the metric.
\begin{equation}
\label{invarMetric}
{g^{\mu \nu }}\xrightarrow[g]{}{g^{\mu \nu }}
\end{equation}
The gauge transformation may be regarded as part of the definition of the wave function, in a way that will be made clear.

The gauge transformation is limited to unitary to conform to state representation in quantum mechanics, which is that of left and right vectors in a complex space. The ``covariant wave function'' here is a ``right vector'', and the contravariant row matrix is the left vector.\cite{SchwingP1} Inner products of a contravariant and covariant matrix (wave function) are \textit{defined} to be gauge invariant. A contravariant row matrix, left vector, wave function ${\psi }$ is \textit{defined} to transform as
\begin{equation}
  {\bm\psi }\xrightarrow[g]{}{\hat {\bm\psi}  = \bm\psi }{{\mathbf{T}}^{ - 1}}
\end{equation}

Limit the gauge transformations to those of interest in quantum mechanics, where the \textit{norm} of the wave function is required to be gauge invariant,
\begin{equation}
\label{norm}
{{\bm\phi }^\dag }{\bm\phi  = }\sum\limits_i {{\phi }_i^*{{\phi }_i}} \xrightarrow[g]{}\widehat {{{\bm\phi }^\dag }{\bm\phi }} = {{\bm\phi }^\dag }{\bm\phi }
\end{equation}
where ${{\bm\phi }^\dag }$ is the conjugate transpose of the wave function. The gauge \textit{contravariant} transformation required to assure invariance of the real-valued, positive definite norm is the conjugate transpose of Eq.~(\ref{gaugeTransPhi}),
\begin{equation}
  \label{contraTransf}
  {{\bm\phi }^\dag }\xrightarrow[g]{}\widehat {{{\bm\phi }^\dag }}{ = }{{\bm\phi }^\dag }{{\mathbf{T}}^\dag }
\end{equation}
Substitute into Eq.~(\ref{norm}).
\begin{equation}
  \label{invarNorm}
{{\bm\phi }^\dag }{\bm\phi }\xrightarrow[g]{}{{\bm\phi }^\dag }{{\mathbf{T}}^\dag }{\mathbf{T}}{\bm\phi }
\end{equation}
Invariance of the norm is achieved when the gauge transformation is limited to \textit{unitary} gauge transformations where,
\begin{equation}
  \label{unitaryTransfMat}
  {{\mathbf{T}}^\dag }{\mathbf{T}} = {\mathbf{T}}{{\mathbf{T}}^\dag } = {\mathbf{1}}
\end{equation}
The gauge transformations examined here are unitary transformations. Under these transformations, any column matrix covariant wave function can be ``converted'' to a contravariant row matrix with the conjugate transpose operation, to make an invariant inner product. The contravariant transformation is,
\begin{equation}
  {\bm\psi }\xrightarrow[g]{}{\hat {\bm\psi }  = \bm\psi }{{\mathbf{T}}^\dag }
\end{equation}

For completeness, the gauge transformation of the square matrix ${\mathbf{M}}$ is defined so that the following product is invariant,
\begin{equation}
  \label{MatInvar}
{{\bm\phi }^\dag }{\mathbf{M}}{\bm\phi }\xrightarrow[g]{}{{\bm\phi }^\dag }{\mathbf{M}}{\bm\phi  = }{{\bm\phi }^\dag }{{\mathbf{T}}^\dag }{\mathbf{\hat MT}}{\bm\phi }
\end{equation}
which implies the transformation properties for the matrix\cite{RamondP189},
\begin{equation}
  \label{MatTrans}
{\mathbf{M}}\xrightarrow[g]{}{\mathbf{\hat M}} = {\mathbf{TM}}{{\mathbf{T}}^\dag }
\end{equation}

We started with the definition of a covariant transformation in Eq.~(\ref{gaugeTransPhi}) as applied to a column matrix wave function, then continued to attach the specific gauge transformation property to specific matrix shapes,
\begin{itemize}
	\item column, covariant, Eq.~(\ref{gaugeTransPhi}),
	\item row, contravariant, Eq.~(\ref{contraTransf}),
	\item square matrix, Eq.~(\ref{MatTrans}),
	\item an invariant of any shape, Eq.~(\ref{invarMetric}).
\end{itemize}
We will maintain this correspondence between the shape of the complex matrix and its gauge transformation properties.

The wave function used here is a column matrix of complex-valued scalars, but several other models for wave functions are used in quantum mechanics. The Dirac equation for the spinor wave function of an electron may taken to be classical Grassmann numbers which anticommute among themselves.\cite{DiracGrass} The electron wave function becomes an operator in a number representation in its interactions with an electromagnetic field.\cite{PeskinSec2P3}

Invariant Lagrangians are typically constructed as a sum of \textit{gauge invariant} terms, and now we have the means to create such terms, such as the product between a contravariant and covariant wave function.

\section{Transformation properties of the gauge covariant derivative}
The \textit{gauge covariant derivative} of the wave function is \textit{defined} to transform the same way as the wave function. Indicate the gauge transformation as applied to the wave function and its derivative as,
\begin{equation}
\label{gaugeTransDPhi}
{D_\mu }{\bm\phi }\xrightarrow[g]{}\widehat {{D_\mu }{\bm\phi }}{ = }{\mathbf{T}}\left( {{D_\mu }{\bm\phi }} \right)
\end{equation}
Put another way, the gauge covariant derivative does not change the transformation properties of its operand, which in this case is the wave function. The gauge transformations also applies to the operator form of the gauge covariant derivative ${D_\mu }$.
\begin{equation}
\label{OperGTrans}
  {D_\mu }{\bm\phi }\xrightarrow[g]{}{\hat D_\mu }{\hat {\bm\phi}  = }{\hat D_\mu }{\mathbf{T}}{\bm\phi }
\end{equation}
Equate the two transformed derivatives on the right hand sides of Eqs~(\ref{gaugeTransDPhi}) and (\ref{OperGTrans}).
\begin{equation}
  {\mathbf{T}}{D_\mu }{\bm\phi  = }{\hat D_\mu }{\mathbf{T}}{\bm\phi }
\end{equation}

The wave function is arbitrary, and solve for the transformed gauge covariant derivative operator using Eq.~(\ref{unitaryTransfMat}).\cite{RamondP192}
\begin{equation}
\label{DoperatorGTrans}
  {D_\mu }\xrightarrow[g]{}{\hat D_\mu }{ = }{\mathbf{T}}{D_\mu }{{\mathbf{T}}^\dag }
\end{equation}
This is the same gauge covariant property as a square matrix defined in Eq.~(\ref{MatTrans}).

The gauge covariant derivative is \textit{defined} to reduce to the ordinary partial derivative when applied to an invariant scalar object. Apply the gauge covariant derivative to an invariant such as the inner product between a contravariant wave function ${{\bm\psi} ^\dag }$ , and a covariant wave function, ${\bm\phi }$,\cite{RamondP192}
\begin{equation}
\label{partialScalar}
  {D_\mu }\left( {{\bm\psi ^\dag }{\bm\phi }} \right){ = }{\partial _\mu }\left( {{{\bm\psi} ^\dag }{\bm\phi }} \right)
\end{equation}
Apply the product rule.
\begin{equation}
\label{productRule}
  {D_\mu }\left( {{{\bm\psi} ^\dag }} \right){\bm\phi } + {{\bm\psi} ^\dag }{D_\mu }\left( {\bm\phi } \right){ = }{\partial _\mu }\left( {{{\bm\psi} ^\dag }{\bm\phi }} \right)
  \end{equation}
 Create another expression by taking the conjugate transpose of the invariant, again resulting in an invariant. Apply the gauge covariant derivative.
\begin{equation}
  {D_\mu }\left( {{{\bm\phi }^\dag }\bm\psi } \right){ = }{\partial _\mu }\left( {{{\bm\phi }^\dag }\bm\psi } \right)
\end{equation}
and the product rule again applies.
\begin{equation}
 \label{productRuleB}
 {D_\mu }\left( {{{\bm\phi }^\dag }} \right)\bm\psi  + {{\bm\phi }^\dag }{D_\mu }\left( \bm\psi  \right){ = }{\partial _\mu }\left( {{{\bm\phi }^\dag }\bm\psi } \right)
\end{equation}
Take the conjugate transpose of Eq.~(\ref{productRule}).
\begin{equation}
\label{ctranProduct}
 {{\bm\phi }^\dag }{D_\mu }\left( \bm\psi  \right) + {\left( {{D_\mu }\left( {\bm\phi } \right)} \right)^\dag }\bm\psi { = }{\partial _\mu }\left( {{{\bm\phi }^\dag }\bm\psi } \right)
\end{equation}
Equate the left hand sides of Eqs.~(\ref{productRuleB}) and (\ref{ctranProduct}).
\begin{equation}
\label{ctDcommutes}
  {\left( {{D_\mu }\left( {\bm\phi } \right)} \right)^\dag }{ = }{D_\mu }\left( {{{\bm\phi }^\dag }} \right)
\end{equation}
The conjugate transpose commutes with the gauge covariant derivative. Removing the arbitrary wave function, and again viewing the gauge covariant derivative as an operator, the operator must be Hermitian. This property arises from interpreting the conjugate transpose as having contravariant gauge transformation properties.
\begin{equation}
\label{DHermitian}
  {\left( {{D_\mu }} \right)^\dag }{ = }{D_\mu }
\end{equation}

Sufficient properties are defined here to begin constructing the gauge covariant derivative.

\section{Construction of the gauge covariant derivative}
The definition of the gauge covariant derivative requires the introduction of an additional gauge field, ${{\mathbf{A}}_\mu }\left( x \right)$, also a square matrix of the same dimension as the transformation matrix. Following the coordinate covariant derivative analogy, the gauge covariant derivative of the wave function  ${\bm\phi }$ is defined    the usual way by including gauge fields by the method of ``minimum coupling''.\cite{theGCD}~ \cite{RamondGCD}~ \cite{AitchisonP25}~ \cite{FelsagerP71} Later we find that this field is an external field applying a force to the wave function system. In the context of electromagnetics, this new field would be the magnetic potential ${A_\mu }$.\cite{HammondP601} Try the following form for the gauge covariant derivative operator as a generalization of the ordinary partial derivative.\cite{RamondP192_12}~ \cite{PeskinP193}
\begin{equation}
\label{gaugeCovMatrixForm}
{D_\mu } = {\mathbf{1}}{\partial _\mu } - i{{\mathbf{A}}_\mu }
\end{equation}
All parts of this \textit{Ansatz} are invariant except for the unknown transformation properties of the new field ${{\mathbf{A}}_\mu }$. Substitute into Eq.~(\ref{DoperatorGTrans}).
\begin{equation}
\label{nonCovTransfA}
{\partial _\mu }{\mathbf{1}} - i{{\mathbf{\hat A}}_\mu }{ = }{\mathbf{T}}\left( {{\mathbf{1}}{\partial _\mu } - i{{\mathbf{A}}_\mu }} \right){{\mathbf{T}}^\dag }
\end{equation}
Solve for the transformed field.\cite{RamondP192_25}
\begin{equation}
\label{Agtran}
{{\mathbf{A}}_\mu }\xrightarrow[g]{}{{\mathbf{\hat A}}_\mu }{ = }i{\mathbf{T}}{\partial _\mu }{{\mathbf{T}}^\dag } + {\mathbf{T}}{{\mathbf{A}}_\mu }{{\mathbf{T}}^\dag }
\end{equation}
The new field is a square matrix, but compare to Eq.~(\ref{MatTrans}) to find an additional term which breaks the linear homogeneous transformation properties of a square matrix.

With the new field transforming as indicated, the gauge covariant derivative defined in Eq.~(\ref{nonCovTransfA}) satisfies the transformation requirements in Eq.~(\ref{partialScalar}). The gauge covariant derivative looks like the following when applied to the wave function.
\begin{equation}
\label{GCDwavefunction}
  {D_\mu }{\bm\phi } = {\partial _\mu }{\bm\phi } - i{{\mathbf{A}}_\mu }{\bm\phi }
\end{equation}
This form of the derivative satisfies the required transformation properties of the gauge covariant derivative.

If there is a field we can add to ${{\mathbf{A}}_\mu }$ indicated by ${{\mathbf{A}}_{ \bot \mu }}$ which \textit{commutes} with the transformation matrix ${\mathbf{T}}$, then Eq.~(\ref{Agtran}) indicates
\begin{equation}
  \left( {{{\mathbf{A}}_\mu } + {{\mathbf{A}}_{ \bot \mu }}} \right)\xrightarrow[g]{}{\mathbf{T}}\left( {{{\mathbf{A}}_\mu } + {{\mathbf{A}}_{ \bot \mu }}} \right){{\mathbf{T}}^{ - 1}} + \left( {{\partial _\mu }{\mathbf{T}}} \right){{\mathbf{T}}^{ - 1}}
\end{equation}
then ${{\mathbf{A}}_{ \bot \mu }}$ is invariant under the gauge transformation.
\begin{equation}
  {{\mathbf{A}}_{ \bot \mu }}\xrightarrow[g]{}{{\mathbf{A}}_{ \bot \mu }}
\end{equation}
Other commuting transformations do not enter the construction of the gauge covariant derivative.

The gauge covariant derivative may also be applied to a contravariant (row matrix) wave function ${\bm\psi ^\dag }$. Substitute the construction into the product rule, Eq.~(\ref{productRule}), then solve for the derivative as applied to a contravariant wave function.
\begin{equation}
  {D_\mu }\left( {{\bm\psi ^\dag }} \right){\bm\phi } + {\bm\psi ^\dag }\left( {{\partial _\mu }{\bm\phi } - i{{\mathbf{A}}_\mu }{\bm\phi }} \right){ = }{\partial _\mu }\left( {{\bm\psi ^\dag }{\bm\phi }} \right)
\end{equation}
Eliminate the arbitrary covariant wave function with the result,
\begin{equation}
\label{contraGCD}
  {D_\mu }{\bm\psi ^\dag }{ = }{\partial _\mu }{\bm\psi ^\dag } + i{\bm\psi ^\dag }{{\mathbf{A}}_\mu }
\end{equation}
Further, the gauge covariant derivative may be applied to a square matrix. All that matters to the construction of the gauge covariant derivative is the gauge transformation properties of the operand, so create a square matrix by using the outer product.
\begin{equation}
  {D_\mu }\left( {{\bm\phi }{\bm\psi ^\dag }} \right){ = \phi }{\partial _\mu }{\bm\psi ^\dag } + i{\bm\phi }{\bm\psi ^\dag }{{\mathbf{A}}_\mu } + \left( {{\partial _\mu }{\bm\phi }} \right){\bm\psi ^\dag } - i{{\mathbf{A}}_\mu }{\bm\phi }{\bm\psi ^\dag }
\end{equation}
or
\begin{equation}
  {D_\mu }\left( {{\bm\phi }{\bm\psi ^\dag }} \right){ = }{\partial _\mu }\left( {{\bm\phi }{\bm\psi ^\dag }} \right) + i\left[ {\left( {{\bm\phi }{\bm\psi ^\dag }} \right),{{\mathbf{A}}_\mu }} \right]
\end{equation}
This can be generalized to an arbitrary square matrix ${\mathbf{M}}$,
\begin{equation}
\label{DerivMat}
  {D_\mu }{\mathbf{M}}{ = }{\partial _\mu }{\mathbf{M}} + i\left[ {{\mathbf{M}},{{\mathbf{A}}_\mu }} \right]
\end{equation}
which transforms as in Eq.~(\ref{MatTrans}).

The gauge covariant derivative operator is Hermition as found in Eq.~(\ref{DHermitian}) which provides an additional property of the new field ${{\mathbf{A}}_\mu }$.
\begin{equation}
  {\left( {{\mathbf{1}}{\partial _\mu } - i{{\mathbf{A}}_\mu }} \right)^\dag } = {\mathbf{1}}{\partial _\mu } - i{{\mathbf{A}}_\mu }
\end{equation}
which implies that the new field is Hermitian.
\begin{equation}
  {{\mathbf{A}}_\mu }{ = }{\mathbf{A}}_\mu ^\dag 
\end{equation}

Essential results come from introducing additional dependency among the elements of the matrix. To provide for further dependency, the transformation matrix may be  \textit{parameterized}.

\section{Parameterization of the transformation}

It becomes extremely useful to assume that the transformation matrix ${\mathbf{T}}\left( \theta  \right)$ is a function    of $n$ independent, real-valued, \textit{parameters} ${\theta ^a}$, $a = 1,2,3 \ldots n$.  The parameters are represented in the aggregate as $\theta$. Each element of the square transformation matrix is a function of the parameters. The parameterization of the transformation matrix is fixed for all calculations to follow and are specific to a \textit{Lie group}. Not all the matrix elements are independent since the number of parameters is assumed fewer than the number of elements in the matrix.

Each of the infinite set of parameterized transformation matrices is uniquely defined by the parameters, so that the parameters   are the \textit{coordinates} of the transformation matrix ${\mathbf{T}}\left( \theta  \right)$  on the $n$ dimensional manifold formed by the parameters. A specific transformation matrix ${\mathbf{T}}\left( {{\theta _1}} \right)$ is located at ${\theta _1}$ on the manifold.

Space-time dependency of the transformation matrix enters only through the parameters so that the full dependency is indicated by ${\mathbf{T}}\left( {\theta \left( x \right)} \right)$. Position in space-time does not explicitly appear in the transformation matrix. Space-time dependency of the transformation enters in a smooth differentiable way only through the parameters ${\theta ^a}\left( x \right)$.

Parameterization of the transformation matrix provides unique identification of the matrices as elements of a Lie group with the addition of the group axioms.\cite{Symmetries} The group axioms each make physical sense as a model of gauge transformations. For example, successive transformations of a wave function again transforms according to a transformation matrix from the same Lie group.

We proceed to the \textit{Lie algebra}, which is a linearization of the transformation matrix near the identity element. However, little reference will be made to group properties, since the principle of covariance provides sufficient guidance for the construction of the gauge covariant derivative.

\section{Taylor expansion of the transformation}
Expand the transformation matrix near the origin in a Taylor series. Without loss of generality, assume that the identity resides at the origin of the coordinates ${\theta ^a} = 0$.
\begin{equation}
  {\mathbf{T}}\left( 0 \right) = {\mathbf{1}}
\end{equation}

Assume that the transformations are connected to the identity by a smooth differentiable path in parameter space. The continuity requirement allows us to take derivatives of the transformation matrix with respect to the parameters. The Taylor series is\cite{WeinbergP54}
\begin{equation}
  {\mathbf{T}} = {\mathbf{T}}\left( 0 \right) + {\left. {\frac{{\partial {\mathbf{T}}\left( \theta  \right)}}{{\partial {\theta ^a}}}} \right|_{\theta  = 0}}{\theta ^a} + O\left( {{\theta ^2}} \right)
\end{equation}
 Fix the point of evaluation at the origin, ${\theta ^a} = 0$, then define the square matrix constant \textit{generators} ${{\mathbf{t}}_a}$  where the imaginary $i$ conveniently connects to a convention made apparent later.
\begin{equation}
\label{defGen}
  i{{\mathbf{t}}_a} = {\left. {\frac{{\partial {\mathbf{T}}\left( \theta  \right)}}{{\partial {\theta ^a}}}} \right|_{\theta  = 0}}
\end{equation}
Approximate the transformation near the origin. Since the transformation matrix must be close to the origin in this approximation, the start and the end points of the parameter difference $\delta {\theta ^a}$ must also be close to the origin.
\begin{equation}
\label{1stOrder}
  {\mathbf{T}}\left( {\delta \theta } \right) \approx {\mathbf{1}} + i{{\mathbf{t}}_a}\delta {\theta ^a}
\end{equation}
The square matrix generators are summed via the repeating index $a$  with a square matrix result,
\begin{equation}
  {{\mathbf{t}}_a}\delta {\theta ^a} = \sum\limits_{a = 1}^N {{{\mathbf{t}}_a}\delta {\theta ^a}} 
\end{equation}
Find that in order to limit our transformations to unitary in Eq.~(\ref{unitaryTransfMat}),
\begin{equation}
  {\left( {{\mathbf{1}} + i{{\mathbf{t}}_a}\delta {\theta ^a}} \right)^\dag }\left( {{\mathbf{1}} + i{{\mathbf{t}}_a}\delta {\theta ^a}} \right) = {\mathbf{1}}
\end{equation}
the generators must be Hermitian.
\begin{equation}
  {\mathbf{t}}_a^\dag  = {{\mathbf{t}}_a}
\end{equation}

The definition of the generators, Eq.~(\ref{defGen}) implies a number of properties. The  generators have no parameter dependence, so  by construction are gauge invariant.
\begin{equation}
\label{invarGen}
  {{\mathbf{t}}_a}\xrightarrow[g]{}{{\mathbf{t}}_a}
\end{equation}

Again by construction the generators are are not a function of position, therefore constant.
\begin{equation}
\label{constGen}
  {\partial _\mu }{{\mathbf{t}}_a} = 0
\end{equation}
By \textit{definition} the gauge covariant derivative of a gauge invariant (and coordinate invariant) object reduces to the ordinary partial derivative, so the partial derivative in Eq.~(\ref{constGen}) can be promoted to a gauge covariant derivative.
\begin{equation}
\label{vanishDerivGen}
{D_\mu }{{\mathbf{t}}_a} = 0
\end{equation}
The vanishing derivative of the generator has the additional advantage of preventing the generators from becoming dynamical objects, and it is an essential property in the derivation of the charged current and its vanishing divergence which follows.

It remains to define the construction of the gauge covariant derivative later. Much can be done using its properties without knowledge of its construction.

\section{Representation of the gauge transformation matrix}
The approximation to the gauge transformation matrix in Eq.~(\ref{1stOrder}) may be repeated in a  limiting process to calculate the matrix located at finite parameter values, so extending the representation to finite distances from the origin.\cite{GilmoreP58}
\begin{equation}
  {\mathbf{T}}\left( \theta  \right) = \mathop {\lim }\limits_{k \to \infty } {\left( {{\mathbf{1}} + i\tfrac{1}{k}{{\mathbf{t}}_a}\delta {\theta ^a}} \right)^k} = \sum\limits_{n = 0}^\infty  {\frac{{{{\left( {i{{\mathbf{t}}_a}{\theta ^a}} \right)}^n}}}{{n!}}} 
\end{equation}
or, using the exponentiation operator,
\begin{equation}
  {\mathbf{T}}\left( \theta  \right) = \operatorname{EXP} \left( {i{{\mathbf{t}}_a}{\theta ^a}} \right)
\end{equation}
This result is valid for real and complex numbers as well as square matrices.

The transformation matrix yielded by this process is constrained by continuity requirements for the Taylor series, hence this process may not yield the transformation matrices in parts of the manifold not smoothly connected to the origin.

\section{Infinitesimal gauge transformation}
Our calculations will use the properties of the wave function under infinitesimal gauge variations, so we will explore the properties of the Lie algebra.

The gauge transformation effectively adds parameter dependence to the transformed wave equation ${\hat {\bm\phi} }\left( {\theta ,x} \right)$. Hence the partial derivative with respect to the parameter of the gauge transformed wave function is,
\begin{equation}
\label{gtransAsDerive}
  \tfrac{\partial }{{\partial {\theta ^a}}}{\hat {\bm\phi} } = \tfrac{\partial }{{\partial {\theta ^a}}}\left( {\mathbf{T}} \right){\bm\phi }
\end{equation}
where the partial with respect to the parameter applies only to the transformation matrix which has the only appearance of the parameter. The partial with respect to the parameter of the wave function and evaluated at the origin is now,
\begin{equation}
  {\left. {\tfrac{\partial }{{\partial {\theta ^a}}}{\hat {\bm\phi} }} \right|_{\theta  = 0}} = i{{\mathbf{t}}_a}{\bm\phi }
\end{equation}

Motivated by use of the partial derivative in Eq.~(\ref{gtransAsDerive}), define a difference or variation operator $\delta$ which is in turn defined by the infinitesimal gauge transformation. The variation operator applies to the wave function just as the partial derivative, but with the evaluation at parameter zero ``built in'' to the notation.
\begin{equation}
  \left( {\tfrac{\delta }{{\delta {\theta ^a}}}} \right){\bm\phi } \equiv {\left. {\tfrac{\partial }{{\partial {\theta ^a}}}{\hat {\bm\phi} }} \right|_{\theta  = 0}} = i{{\mathbf{t}}_a}{\bm\phi }
\end{equation}

Substitute Eq.~(\ref{1stOrder}) into the gauge transformation of the wave function Eq.~(\ref{gaugeTransDPhi}), so that for parameter values infinitesimally near the origin,  $\delta {\theta ^a}$, defines a change indicated by $\delta$,
\begin{equation}
{\hat {\bm\phi}  = \bm\phi  + }\delta {\bm\phi  = }\left( {{\mathbf{1}} + i{{\mathbf{t}}_a}\delta {\theta ^a}} \right){\bm\phi }
\end{equation}
so that the infinitesimal gauge transformation is,
\begin{equation}
\label{theInfGaugeTransf}
  \begin{gathered}
  \delta {\bm\phi } = \left( {i{{\mathbf{t}}_a}{\bm\phi }} \right)\delta {\theta ^a} \\ 
  \delta \left( {{D_\mu }{\bm\phi }} \right) = \left( {i{{\mathbf{t}}_a}{D_\mu }{\bm\phi }} \right)\delta {\theta ^a} \\ 
\end{gathered} 
\end{equation}
or in the form of a derivative,
\begin{equation}
\label{gaugeTranofwaveF}
  \begin{gathered}
  \left( {\tfrac{\delta }{{\delta {\theta ^a}}}} \right){\bm\phi } = i{{\mathbf{t}}_a}{\bm\phi } \\ 
  \left( {\tfrac{\delta }{{\delta {\theta ^a}}}} \right)\left( {{D_\mu }{\bm\phi }} \right) = i{{\mathbf{t}}_a}\left( {{D_\mu }{\bm\phi }} \right) \\ 
\end{gathered} 
\end{equation}

Contravariant objects have the transformation property,
\begin{equation}
  \begin{gathered}
  \delta {{\bm\phi }^\dag } = \left( { - i{{\bm\phi }^\dag }{{\mathbf{t}}_a}} \right)\delta {\theta ^a} \\ 
  \delta \left( {{D_\mu }{{\bm\phi }^\dag }} \right) = \left( { - i\left( {{D_\mu }{{\bm\phi }^\dag }} \right){{\mathbf{t}}_a}} \right)\delta {\theta ^a} \\ 
\end{gathered} 
\end{equation}
Substitute Eq.~(\ref{1stOrder}) into Eq.~(\ref{MatTrans}) to find the  infinitesimal gauge transformation for a square matrix.
\begin{equation}
\label{infGTranM}
  \delta {{\rm M} = }i\left[ {{{\mathbf{t}}_a},{{\rm M}}} \right]\delta {\theta ^a}
 \end{equation}

Again, the generators are gauge invariant, according to Eq.~(\ref{invarGen}).
\begin{equation}
\label{diffInvarGen}
  \delta {{\mathbf{t}}_a} = 0
\end{equation}
and the metric is gauge invariant,
\begin{equation}
  \delta {g^{\mu \nu }} = 0
\end{equation}

The infinitesimal gauge transformation of ${{\mathbf{A}}_\mu }$ follows from substituting Eq.~(\ref{1stOrder}) into the transformation Eq.~(\ref{Agtran}).
\begin{equation}
  {{\mathbf{\hat A}}_\mu }{ = }i\left( {{\mathbf{1}} + i{{\mathbf{t}}_a}\delta {\theta ^a}} \right){\partial _\mu }\left( {{\mathbf{1}} - i{{\mathbf{t}}_b}\delta {\theta ^b}} \right) + \left( {{\mathbf{1}} + i{{\mathbf{t}}_a}\delta {\theta ^a}} \right){{\mathbf{A}}_\mu }\left( {{\mathbf{1}} - i{{\mathbf{t}}_b}\delta {\theta ^b}} \right)
\end{equation}
so that,
\begin{equation}
  {{\mathbf{\hat A}}_\mu }{ = }{{\mathbf{t}}_b}{\partial _\mu }\delta {\theta ^b} + {{\mathbf{A}}_\mu } + i\left[ {{{\mathbf{t}}_a},{{\mathbf{A}}_\mu }} \right]\delta {\theta ^a}
\end{equation}
or
\begin{equation}
\label{AgaugeTransf}
  \delta {{\mathbf{A}}_\mu }{ = }{{\mathbf{t}}_b}{\partial _\mu }\delta {\theta ^b} + i\left[ {{{\mathbf{t}}_a},{{\mathbf{A}}_\mu }} \right]\delta {\theta ^a}
\end{equation}

This completes the application of the infinitesimal gauge transformation to the wave function, and its derivative, and generators of the gauge transformation. However, the toolkit required to construct gauge invariant Lagrangians is not complete, since terms in the Lagrangian may include the parameter index. The gauge transformation properties of column, row and square matrices are defined above, but remaining to be defined is gauge transformation properties of the parameter index of the generators. The definition of gauge invariance of the generators will lead to Sophus Lie's second theorem, and consistent infinitesimal transformation properties of the parameter index.

\section{Gauge invariant generators, Sophus Lie's second theorem}
The generators are defined to be gauge invariant, then the statement of that invariance,  Eq.~(\ref{diffInvarGen}) links two infinitesimal homogeneous gauge transformations, one for the matrix indices as already discussed, and the other for the parameter index, which has not been examined. Exactly analogous is the simultaneous gauge transformation of the wave function and electromagnetic field as outlined in Eqs.~(\ref{electroPhase}) and (\ref{electroGauge}).

Each generator is a square matrix so that the matrix part of the infinitesimal gauge transformation must look like Eq.~(\ref{infGTranM}). Now we turn attention to how a parameter index is transformed under a gauge transformation.

The generators ${{\mathbf{t}}_a}$ form a basis for a vector space. One is then free to redefine the generators by a real nonsingular linear transformation with a corresponding redefinition of the parameters\cite{SchwingLinSp}. The required invariance of these generators as stated in  Eq.~(\ref{diffInvarGen}) dictates that the parameters must transform in such a way as to preserve invariance under infinitesimal variations of the parameters. These infinitesimal variations, because of their smallness, may be expected to effect a homogeneous transformation of the generators.

We have defined the infinitesimal gauge transformation for column, row and square matrices. Now to find the infinitesimal gauge transformation of the \textit{parameter} index. With the left and right multiplications of wave functions, the only free index in the expression ${{\bm\psi }^\dag }{{\mathbf{t}}_a}{\bm\phi }$ is the covariant \textit{parameter} index ``$a$''. This parameter index must transform under a gauge transformation in order to satisfy the requirement that the generators are gauge invariant.

In the same spirit as the homogeneous transformation of the wave function in Eq.~(\ref{gaugeTransPhi}), we \textit{define} the most general homogeneous transformation in terms of the set of constants $f_{\;\;ba}^c$ which will turn out to be the structure constants of the \textit{Lie algebra}. Use the structure constants to define the following transformation for the covariant (lower) parameter index.
\begin{equation}
  \left( {\delta /\delta {\theta ^b}} \right)\left( {{{\bm\psi }^\dag }{{\mathbf{t}}_a}{\bm\phi }} \right) = f_{\;\;ba}^c\left( {{{\bm\psi }^\dag }{{\mathbf{t}}_c}{\bm\phi }} \right)
\end{equation}
Generalize this infinitesimal transformation property to any lowered parameter index such as an arbitrary complex-valued vector ${V_a}\left( \theta  \right)$.
\begin{equation}
\label{covarParamTransf}
  \left( {\tfrac{\delta }{{\delta {\theta ^b}}}} \right){V_a}{ = }f_{\;\;ba}^c{V_c}
\end{equation}
An inner product with a \textit{contravariant} parameter vector ${W^a}$ is invariant.
\begin{equation}
  \delta \left( {{V_a}{W^a}} \right){ = }0
\end{equation}
Once again, this invariance requirement leads to the infinitesimal gauge transformation for a contravariant parameter vector (raised index).
\begin{equation}
\label{contraParamTransf}
  \left( {\tfrac{\delta }{{\delta {\theta ^b}}}} \right){V^a} =  - f_{\;\;bc}^a{V^c}
\end{equation}
A procedure exists to \textit{convert} a covariant to a contravariant vector via the \textit{Cartan-Killing metric} which will be described shortly. The procedure is analogous to raising and lowering coordinate tensor indices with the metric tensor, and will be defined shortly.

We have sufficient definitions to elaborate the infinitesimal gauge invariance of the generators as stated in Eq.~(\ref{invarGen}). The requirement for invariant generators links the two homogeneous gauge transformations, one for the square matrix generators, Eq.~(\ref{infGTranM}), and the other for the lowered parameter index Eq.~(\ref{covarParamTransf}).
\begin{equation}
\tfrac{\delta }{{\delta {\theta ^b}}}\left( {{{\mathbf{t}}_a}} \right){ = }f_{\;\;ba}^c{{\mathbf{t}}_c} + i\left[ {{{\mathbf{t}}_b},{{\mathbf{t}}_a}} \right] \equiv 0
\end{equation}
The generators are defined to be invariant under an infinitesimal gauge transformation, so that this expression vanishes. We see that the infinitesimal gauge transformation applied to the matrix generators, is compensated by a redefinition of the parameter for each generator, so that the net result is zero:  invariant generators.  Rewrite this expression.\cite{WeinP54} 
\begin{equation}
\label{Lie2ndTh}
  \left[ {{{\mathbf{t}}_a},{{\mathbf{t}}_b}} \right] = if_{\;\;ab}^c{{\mathbf{t}}_c}
\end{equation}
where the structure constants are \textit{defined} to be antisymmetric.
\begin{equation}
\label{SophusLie}
  f_{\;\;ab}^c =  - f_{\;\;ba}^c
\end{equation}
Eq.~{\ref{SophusLie}) is Lie's Second Theorem\cite{Gilmore1974P100} (Marius Sophus Lie 1842 – 1899), which we will refer to as closure under the operation of commutation.\cite{Gilmore2006P61} This theorem follows from the requirement that both matrix and parameter indices transform homogeneously in such a way as to preserve the infinitesimal gauge invariance of the generators. As pointed out in the references, imposition of group axioms on the gauge transformation matrix leads to the same destination.

\section{Vanishing derivative of the generators}
The gauge covariant derivative of the generators vanish in Eq.~(\ref{vanishDerivGen}). This must be confirmed with the explicit definition of the derivative in Eq.~(\ref{GCDwavefunction}). The gauge covariant derivative must be extended to include parameter indices as well as matrix indices. Exploit the deep analogy\cite{WeinDeepAnalogy} to the coordinate covariant derivative by adding a term, $\Gamma _{\mu a}^c$, analogous to the Christoffel symbol appearing in the coordinate covariant derivative.\cite{CarmeliP53}
\begin{equation}
\label{DTZero}
  {D_\mu }{{\mathbf{t}}_a} = {\partial _\mu }{{\mathbf{t}}_a} + \Gamma _{\mu a}^c{{\mathbf{t}}_c}{ - }i\left[ {{{\mathbf{A}}_\mu },{{\mathbf{t}}_a}} \right] = 0
\end{equation}
\textit{Assume} that the additional field, ${{\mathbf{A}}_\mu }\left( x \right)$ is within the span of the generators used as basis functions. \cite{RamondP195E13}}~ \cite{CarmeliP9}  
\begin{equation}
\label{Arep}
  {{\mathbf{A}}_\mu } = A_\mu ^a{{\mathbf{t}}_a}
\end{equation}
where the coefficients $A_\mu ^a\left( x \right)$ of the basis functions carry the position dependency. These coefficients are the multicomponented \textit{gauge potential}. Since ${{\mathbf{A}}_\mu }$ is Hermitian, Eq.~(\ref{AHermitian}), and the generators are Hermitian, then the gauge potential is real-valued which fits the convention for the electromagnetic 4-potential.
\begin{equation}
  {\left( {A_\mu ^a} \right)^*} = A_\mu ^a
\end{equation}

Substitute Eqs.~(\ref{constGen}), and (\ref{Arep}) into Eq.~(\ref{DTZero}).
\begin{equation}
  \Gamma _{\mu a}^c{{\mathbf{t}}_c}{ - }i\left[ {A_\mu ^b{{\mathbf{t}}_b},{{\mathbf{t}}_a}} \right] = 0
\end{equation}
Substitute Lie's Second Theorem, Eq.~(\ref{Lie2ndTh}).
\begin{equation}
\label{step2CSym}
  \Gamma _{\mu a}^c{{\mathbf{t}}_c}{ = }A_\mu ^bf_{\;\;ab}^c{{\mathbf{t}}_c}
\end{equation}
Remove the generator dependency.
\begin{equation}
\label{gaugeConnection}
  \Gamma _{\mu a}^c{ = }A_\mu ^bf_{\;\;ab}^c
\end{equation}
The vanishing covariant derivative of the generators provides a consistent calculation, Eq.~(\ref{Arep}) for the additional field, ${{\mathbf{A}}_\mu }\left( x \right)$. Substitute Eqs.~(\ref{Arep}) and (\ref{gaugeConnection}) into (\ref{DTZero}).
\begin{equation}
\label{vanishingDFull}
  {D_\mu }{{\mathbf{t}}_a} = {\partial _\mu }{{\mathbf{t}}_a} + A_\mu ^bf_{\;\;ab}^c{{\mathbf{t}}_c}{ - }iA_\mu ^b\left[ {{{\mathbf{t}}_b},{{\mathbf{t}}_a}} \right] = 0
\end{equation}

Insert gauge potentials via Eq.~(\ref{Arep}) into the gauge covariant derivative, Eqs.~(\ref{GCDwavefunction}), (\ref{contraGCD}) and (\ref{DerivMat}).
\begin{equation}
\label{GCovderivPhi}
{D_\mu }{\bm\phi } = {\partial _\mu }{\bm\phi } - iA_\mu ^a{{\mathbf{t}}_a}{\bm\phi }
\end{equation}
\begin{equation}
  {D_\mu }{\psi ^\dag }{ = }{\partial _\mu }{\psi ^\dag } + iA_\mu ^a{\psi ^\dag }{{\mathbf{t}}_a}
\end{equation}
\begin{equation}
  {D_\mu }{\mathbf{M}}{ = }{\partial _\mu }{\mathbf{M}} + iA_\mu ^a\left[ {{\mathbf{M}},{{\mathbf{t}}_a}} \right]
\end{equation}

The gauge transformation of the gauge potentials $A_\mu ^a\left( x \right)$ follows from substituting Eq.~(\ref{Arep}) into Eq.~(\ref{AgaugeTransf}).
\begin{equation}
\delta \left( {A_\mu ^b{{\mathbf{t}}_b}} \right){ = }{{\mathbf{t}}_b}{\partial _\mu }\delta {\theta ^b} + i\left[ {{{\mathbf{t}}_a},A_\mu ^b{{\mathbf{t}}_b}} \right]\delta {\theta ^a}
\end{equation}
Extract the gauge potential from the commutator, note that the generators are gauge invariant, then substitute Lie's Second Theorem, Eq.~(\ref{Lie2ndTh}).
\begin{equation}
  \delta \left( {A_\mu ^b} \right){{\mathbf{t}}_b}{ = }{{\mathbf{t}}_b}{\partial _\mu }\delta {\theta ^b} - A_\mu ^cf_{\;\;ac}^b{{\mathbf{t}}_b}\delta {\theta ^a}
\end{equation}
Remove the generator dependency, and rename indices.
\begin{equation}
\label{ApotGtrans}
  \delta A_\mu ^a{ = }{\partial _\mu }\left( {\delta {\theta ^a}} \right) - A_\mu ^cf_{\;\;bc}^a\delta {\theta ^b}
\end{equation}
Interestingly, this relation has no matrix indices, so that it is independent of the matrix representation of the generators.

\section{Vectors in parameter space}

Generalize the gauge covariant derivative to an arbitrary covariant vector ${V_a}\left( x \right)$, and contravariant vector ${V^a}\left( x \right)$ with a parameter index. \textit{Define} the following, consistent with Eq.~(\ref{vanishingDFull}),
\begin{equation}
\label{paramVderiv}
  \begin{gathered}
  {D_\mu }\left( {{V_a}} \right) = {\partial _\mu }\left( {{V_a}} \right) + A_\mu ^bf_{\;\;ab}^c{V_c} \hfill \\
  {D_\mu }\left( {{V^a}} \right) = {\partial _\mu }\left( {{V^a}} \right) - A_\mu ^bf_{\;\;cb}^a{V^c} \hfill \\ 
\end{gathered} 
\end{equation}
so that the derivative of an invariant becomes the ordinary partial.
\begin{equation}
  {D_\mu }\left( {{V_a}{V^a}} \right) = {\partial _\mu }\left( {{V_a}{V^a}} \right)
\end{equation}
The gauge covariant derivative and gauge transformation can now be consistently applied to arbitrary contravariant wave functions and vectors in parameter space.

The gauge potential transformation in Eq.~(\ref{ApotGtrans}) can be seen to be closely related to contravariant parameter transformation when compared to Eq.~(\ref{paramVderiv}). In fact, the gauge transformation can be written in an interesting form in terms of a derivative.\cite{RamondP194}
\begin{equation}
  \delta A_\mu ^a{ = }{D_\mu }\left( {\delta {\theta ^a}} \right)
\end{equation}
The form of the additional field in Eq.~(\ref{Arep}) may be generalized by adding a term which commutes with all the generators. The derivative of the generators still vanish with this term, so this is a consistent modification, the implication of which is not pursued here.

\section{Definition of the field strength tensor}
Needed shortly is the commutator of the gauge covariant derivative which can be evaluated given the definition of the derivative, Eq.~(\ref{GCovderivPhi}), constant generators, and Lie's second theorem Eq.~(\ref{Lie2ndTh}).
\begin{equation}
\label{gcdComm}
  \left[ {{D_\mu },{D_\nu }} \right]{\bm\phi } = iF_{\mu \nu }^a{{\mathbf{t}}_a}{\bm\phi }
\end{equation}
where the \textit{definition} of the field strength tensor $F_{\mu \nu }^a$ for the gauge potentials $A_\mu ^a$ is
\begin{equation}
\label{nonAbFTens}
  F_{\mu \nu }^a =  - {\partial _\mu }A_\nu ^a + {\partial _\nu }A_\mu ^a - A_\mu ^bA_\nu ^cf_{\;\;bc}^a
\end{equation}
The field strength tensor is free of any matrix indices, hence is independent of the matrix representation of the generators. As indicated by the contravariant parameter index the field strength tensor is \textit{not} gauge invariant,
\begin{equation}
  \left( {\tfrac{\delta }{{\delta {\theta ^b}}}} \right)F_{\mu \nu }^a =  - f_{\;\;bc}^aF_{\mu \nu }^c
\end{equation}
whereas the electromagnetic field tensor \textit{is} gauge invariant.

This completes the set of tools required to do gauge covariant calculations. All objects such as generators, wave functions, field strength tensor, current vectors, have well defined transformation properties and derivatives.

\section{Definition of the Cartan-Killing metric}
We use the Cartan-Killing metric ${g_{ab}}$ to raise and lower parameter indices, so ``converting'' one to the other.

Apply the Cartan–Killing \textit{inner product} which is defined as the \textit{trace} of the matrix product of two generators.\cite{GilmoreP65E437}  The curly brackets indicate application of the matrix Trace operation.
\begin{equation}
  {g_{ab}} \equiv Tr\left\{ {{{\mathbf{t}}_a}{{\mathbf{t}}_b}} \right\}
\end{equation}
By this definition, the metric is symmetric since matrices commute under the trace.
\begin{equation}
  {g_{ab}} = {g_{ba}}
\end{equation}

The metric, constructed from gauge invariant generators, is gauge invariant.
\begin{equation}
\tfrac{\delta }{{\delta {\theta ^c}}}\left( {{g_{ab}}} \right) = 0
\end{equation}
Substitute the variation for a covariant parameter vector, Eq.~(\ref{covarParamTransf}).
\begin{equation}
  \tfrac{\delta }{{\delta {\theta ^c}}}\left( {{g_{ab}}} \right) = f_{\;\;ca}^e{g_{eb}} + f_{\;\;cb}^e{g_{ae}} = 0
\end{equation}
Define the \textit{covariant} structure constants,\cite{WeinP9} 
\begin{equation}
  {f_{abc}} = f_{\;\;bc}^d{g_{da}}
\end{equation}
then,
\begin{equation}
  \tfrac{\delta }{{\delta {\theta ^c}}}\left( {{g_{ab}}} \right) = {f_{bca}} + {f_{acb}} = 0
\end{equation}
which shows that the covariant structure constants $f_{bca}$ are antisymmetric in the $ab$ indices.
\begin{equation}
{f_{bca}} =  - {f_{acb}}
\end{equation}
The definition of the covariant structure constants implies antisymmetry in the $ca$ indices, so that they are completely antisymmetric.

Similarly, we expect the gauge covariant derivative of a function of the generators to vanish.
\begin{equation}
  {D_\mu }\left( {{g_{ab}}} \right) = {\partial _\mu }{g_{ab}} + A_\mu ^df_{\;\;ad}^c{g_{cb}} + A_\mu ^df_{\;\;bd}^c{g_{ac}} = 0
\end{equation}
The Cartan–Killing metric is constant, and the external gauge potential arbitrary, so that the derivative vanishes due to the antisymmetry of the covariant structure constants.

Indicate the \textit{inverse} of the Cartan-Killing metric as ${g^{ab}}$ where
\begin{equation}
  {g^{ac}}{g_{bc}} = \delta _b^a
\end{equation}

The metric and its inverse can be used to ``convert'' parameter indices, so that by definition, given a contravariant vector ${V^a}$ the corresponding covariant vector is
\begin{equation}
  {V_b} = {g_{ab}}{V^a}
\end{equation}

The contravariant parameter index field tensor $F_{\mu \nu }^a$ is defined in Eq.~(\ref{nonAbFTens}).  Construct the covariant parameter field tensor using the Cartan-Killing metric,
\begin{equation}
  {F_{a\;\mu \nu }} = {g_{ab}}F_{\mu \nu }^a
\end{equation}
Construct a gauge and coordinate invariant Lagrangian analogous to electromagnetics, where contravariant parameter indices must be summed against a covariant index such as
\begin{equation}
  {L_{gauge}}\left( {F_{\mu \nu }^a} \right) = F_a^{\mu \nu }F_{\mu \nu }^a
\end{equation}
Compare this to the electromagnetic Lagrangian, also gauge and coordinate invariant.
\begin{equation}
  {L_{elect}}\left( {{F^{\kappa \eta }}} \right) = {F^{\kappa \eta }}{F_{\kappa \eta }}
\end{equation}

The problem of constructing contravariant parameter vectors uses the Cartan-Killing Lie algebra metric which may be used to raise and lower parameter indices, analogous to the coordinate metric.

\section{Invariant structure constants, or the Jacobi identity}
We briefly continue our foray into Lie algebra, in the guise of the infinitesimal gauge transformation. Quadratic constraints upon the structure constants $f_{\;\;\;ab}^c$ follow from the Jacobi identity.\cite{RamondP195} Evaluate the \textit{Jacobi identity} which is the commutator of the matrix generators summed with permutations of the indices.
\begin{equation}
  \left[ {\left[ {{{\mathbf{t}}_a},{{\mathbf{t}}_b}} \right],{{\mathbf{t}}_c}} \right] + \left[ {\left[ {{{\mathbf{t}}_b},{{\mathbf{t}}_c}} \right],{{\mathbf{t}}_a}} \right] + \left[ {\left[ {{{\mathbf{t}}_c},{{\mathbf{t}}_a}} \right],{{\mathbf{t}}_b}} \right] = 0
\end{equation}
This expression is identically zero with expansion of the commutators, and application of the associative property for matrix operations. The Jacobi identity becomes a consistency constraint on the structure constants by repeatedly substituting Eq.~(\ref{Lie2ndTh}).
\begin{equation}
  f_{\;\;ab}^df_{\;\;dc}^e{{\mathbf{t}}_e} + f_{\;\;bc}^df_{\;\;da}^e{{\mathbf{t}}_e} + f_{\;\;ca}^df_{\;\;db}^e{{\mathbf{t}}_e} = 0
\end{equation}
Without constraining the generators, the structure constants must satisfy the quadratic constraint,
\begin{equation}
  f_{\;\;ab}^df_{\;\;dc}^e + f_{\;\;bc}^df_{\;\;da}^e + f_{\;\;ca}^df_{\;\;db}^e = 0
\end{equation}
Note that the quadratic constraint implies gauge invariance of the structure constants via Eqs.~(\ref{covarParamTransf}) and (\ref{contraParamTransf}).\cite{BlagoP377}
\begin{equation}
  \left( {\delta /\delta {\theta ^f}} \right)\left( {f_{\;\;ab}^e} \right) = f_{\;\;ab}^df_{\;\;dc}^e + f_{\;\;bc}^df_{\;\;da}^e + f_{\;\;ca}^df_{\;\;db}^e = 0
\end{equation}
View the Jacobi identity as a consistency requirement since it implies invariance of the structure constants, which is also consistent with invariance of the generators.

\section{The gauge transformation of the Lagrangian}
Now consider the two ``divergence'' applications of the gauge covariant derivative mentioned earlier. Both involve the Lagrangian. Consider the gauge transformation of the Lagrangian.
\begin{equation}
  L\left( {{\bm\phi },{D_\mu }{\bm\phi },{g^{\mu \nu }}} \right)\xrightarrow[g]{}\hat L{ = }\hat L\left( {{\mathbf{T}}{\bm\phi },{\mathbf{T}}{D_\mu }{\bm\phi },{g^{\mu \nu }}} \right)
\end{equation}
where $\hat L$ is the gauge transformed Lagrangian with the replacement indicated in Eqs.~(\ref{gaugeTransPhi}), and (\ref{gaugeTransDPhi}). The transformation adds the matrix   at each appearance of the wave function and its derivative. The replacement adds a parameter   dependency to the Lagrangian. Evaluated at parameter zero, the value of the Lagrangian is unchanged since the transformation matrix reduces to the identity matrix.
\begin{equation}
  {\left. {\hat L\left( {{\mathbf{T}}\left( \theta  \right){\bm\phi },{\mathbf{T}}\left( \theta  \right){D_\mu }{\bm\phi },{g^{\mu \nu }}} \right)} \right|_{\theta  = 0}} = L\left( {{\bm\phi },{D_\mu }{\bm\phi },{g^{\mu \nu }}} \right)
\end{equation}
Apply the partial derivative with respect to $\theta$ to $\hat L$. Since the addition of the transformation matrix does not change the functional dependence, we can evaluate the variation using the chain rule with respect to the wave function and its derivative. 
\begin{equation}
  \left( {\tfrac{\delta }{{\delta {\theta ^a}}}} \right)L = \frac{{\partial L}}{{\partial {\bm\phi }}}\left( {\tfrac{\delta }{{\delta {\theta ^a}}}{\bm\phi }} \right) + \frac{{\partial L}}{{\partial {D_\mu }{\bm\phi }}}\left( {\tfrac{\delta }{{\delta {\theta ^a}}}{D_\mu }{\bm\phi }} \right)
\end{equation}
Substitute the variations, Eq.~(\ref{gaugeTranofwaveF}).
\begin{equation}
\label{varOfL}
  \left( {\tfrac{\delta }{{\delta {\theta ^a}}}} \right)L = \frac{{\partial L}}{{\partial {\bm\phi }}}i{{\mathbf{t}}_a}{\bm\phi } + \frac{{\partial L}}{{\partial {D_\mu }{\bm\phi }}}i{{\mathbf{t}}_a}{D_\mu }{\bm\phi }
\end{equation}
This the starting point for determining the divergence of the current as defined by an infinitesimal gauge transformation.

\section{The divergence of the charged current}
The vanishing divergence of the current will be our first ``vanishing divergence'' relation. The \textit{second} vanishing divergence soon to follow applies to the energy-momentum tensor.

The constant generators expressed in Eq.~(\ref{vanishDerivGen}) implies that the generator commutes with the derivative.
\begin{equation}
\label{DTcomm}
  {{\mathbf{t}}_a}{D_\mu }{\bm\phi  = }{D_\mu }\left( {{{\mathbf{t}}_a}{\bm\phi }} \right)
\end{equation}
so that the variation of the Lagrangian Eq.~(\ref{varOfL}) becomes,
\begin{equation}
\label{chainRL}
  \left( {\tfrac{\delta }{{\delta {\theta ^a}}}} \right)L = \frac{{\partial L}}{{\partial {\bm\phi }}}i{{\mathbf{t}}_a}{\bm\phi } + \frac{{\partial L}}{{\partial {D_\mu }{\bm\phi }}}{D_\mu }\left( {i{{\mathbf{t}}_a}{\bm\phi }} \right)
\end{equation}

Rearrange with application of the product rule,
\begin{equation}
\label{varLExpanded}
  \left( {\tfrac{\delta }{{\delta {\theta ^a}}}} \right)L = \left\{ {\frac{{\partial L}}{{\partial {\bm\phi }}} - {D_\mu }\left( {\frac{{\partial L}}{{\partial {D_\mu }{\bm\phi }}}} \right)} \right\}{{\mathbf{t}}_a}{\bm\phi } + {D_\mu }\left( {\frac{{\partial L}}{{\partial {D_\mu }{\bm\phi }}}{{\mathbf{t}}_a}{\bm\phi }} \right)
\end{equation}
Identify the equations of motion for the wave function ${\bm\phi }$ in the curly brackets. The row-matrix expression ${\bm\Lambda }$  is set to zero to arrive at the Euler-Lagrange equation ${\bm\Lambda } = 0$.\cite{CLewis}
\begin{equation}
\label{eulerLagrange}
  {\bm\Lambda  = }\frac{{\partial L}}{{\partial {\bm\phi }}} - {D_\mu }\left( {\frac{{\partial L}}{{\partial {D_\mu }{\bm\phi }}}} \right) = 0
\end{equation}
\begin{quotation}
  Note:  The many examples of wave equations defined by the Euler-Lagrange equation applied to an appropriate Lagrangian include electromagnetics, the Klein-Gordon wave equation for spin zero fields and the Dirac wave equation for spin one-half fields. Satisfaction of the Euler-Lagrange equation, $\bm\Lambda  = 0$, defines the term ``on shell''. Off shell dynamics violate the Euler-Lagrange equation.
\end{quotation}

For a gauge invariant scalar Lagrangian, the partial with respect to the covariant wave function yields a contravariant result. Similarly, construction of ${\bm\Lambda }$ which includes partials with respect to covariant quantities, indicates that it transforms contravariantly, Eq.~(\ref{contraTransf}).
\begin{equation}
  {\bm\Lambda }\xrightarrow[g]{}{\hat {\bm\Lambda}  = \bm\Lambda }{{\mathbf{T}}^\dag }
\end{equation}
Upon setting the contravariant Euler-Lagrange equations to zero as an equation of motion, subsequent gauge transformations will maintain the zero value for $\hat {\bm\Lambda}$, and so maintain the equations of motion.\cite{LovelockP185}

Each parameter of the gauge transformation has a corresponding current as indicated by the covariant (lower) index ``$a$'' in the \textit{definition},
\begin{equation}
\label{current}
  J_a^\mu  = \frac{{\partial L}}{{\partial {D_\mu }{\bm\phi }}}{{\mathbf{t}}_a}{\bm\phi }
\end{equation}
Substitute the Euler-Lagrange expression Eq.~(\ref{eulerLagrange}) and the current into Eq.~(\ref{varLExpanded}),
\begin{equation}
  \left( {\tfrac{\delta }{{\delta {\theta ^a}}}} \right)L = {\bm\Lambda }{{\mathbf{t}}_a}{\bm\phi } + {D_\mu }J_a^\mu 
\end{equation}
Solve for the divergence of the gauge current.
\begin{equation}
\label{divCurrent}
  {D_\mu }J_a^\mu  = \left( {\tfrac{\delta }{{\delta {\theta ^a}}}} \right)L - i{\bm\Lambda }{{\mathbf{t}}_a}{\bm\phi }
\end{equation}
The gauge current has a vanishing divergence with satisfaction of 1) the Euler-Lagrange equation, and 2) a \textit{symmetry} of the Lagrangian which is \textit{defined} as,
\begin{equation}
  \left( {\tfrac{\delta }{{\delta {\theta ^a}}}} \right)L = 0
\end{equation}
A gauge transformation which is also a symmetry of the Lagrangian, leaves the Lagrangian invariant. The divergence of the current in Eq.~(\ref{divCurrent}) may be regarded as an identity since identity since it is dependent upon only definitions of quantities, and the product rule.

The current is defined in Eq.~(\ref{current}) whether or not the gauge transformation is a symmetry of the Lagrangian.

The definition of a charged current, and the conditions for the vanishing divergence of the current is central in physics, and, for electromagnetics, implies charged current continuity. This relationship motivates and rewards our pursuit for a gauge covariant derivative  that commutes with the generators, Eq.~(\ref{DTcomm}).

\section{The Divergence of the energy-momentum tensor}
The energy-momentum of a system responds to external forces acting upon the system according to a ``divergence law'' as applied to energy-momentum. These interesting results can be found without detailing the construction of the gauge covariant derivative. All that is required is the chain rule, product rule, and the non-commuting gauge covariant derivative.

Apply the gauge covariant derivative to the scalar Lagrangian, Eq.~(\ref{Lagrangian}), and use the chain rule.
\begin{equation}
\label{chainR2}
{D_\mu }L = \frac{{\partial L}}{{\partial {\bm\phi }}}{D_\mu }{\bm\phi } + \frac{{\partial L}}{{\partial \left( {{D_\nu }{\bm\phi }} \right)}}{D_\mu }\left( {{D_\nu }{\bm\phi }} \right) + \frac{{\partial L}}{{\partial \left( {{g^{\lambda \nu }}} \right)}}{D_\mu }{g^{\lambda \nu }}
\end{equation}
The last term vanishes by the metricity condition above, Eq.~(\ref{metricity}). Rearrangement using only the product rule results in
\begin{equation}
\label{rearrangeM}
\begin{gathered}
  {D_\nu }\left\{ {L\delta _\mu ^\nu { - }\frac{{\partial L}}{{\partial {D_\nu }{\bm\phi }}}{D_\mu }{\bm\phi }} \right\} = \frac{{\partial L}}{{\partial {D_\nu }{\bm\phi }}}\left[ {{D_\mu },{D_\nu }} \right]{\bm\phi } \\ 
  { + }\left\{ {\frac{{\partial L}}{{\partial {\bm\phi }}}{ - }{D_\nu }\left( {\frac{{\partial L}}{{\partial {D_\nu }{\bm\phi }}}} \right)} \right\}{D_\mu }{\bm\phi } \\ 
\end{gathered} 
\end{equation}
Verify this expression by expanding the curly brackets, and finding that the end result is Eq.~(\ref{chainR2}), the chain rule. Conversely, prove this expression by noting that each step involving the product rule is reversible.

Identify the equations of motion for the wave function $\bm\phi$ provided by the row-matrix Euler-Lagrange equation ${\bm\Lambda } = 0$.

Identify in Eq.~(\ref{rearrangeM}) the canonical energy-momentum tensor (EMT) ${T_{\mu }}^\nu$, which is the covariant generalization of energy and momentum in space-time,\cite{LandauRef2, MichioP27e165}
\begin{equation}
\label{EMT}
{T_{\mu }}^\nu  = L\delta _\mu ^\nu { - }\frac{{\partial L}}{{\partial {D_\nu }{\bm\phi }}}{D_\mu }{\bm\phi }
\end{equation}
The canonical energy-momentum is the covariant generalization of energy and momentum in space-time.\cite{canonicalEMT}  Substitute the EMT and the Euler-Lagrange expression $\bm\Lambda$ into Eq.~(\ref{rearrangeM}) to find the divergence of the EMT. The gauge covariant derivative ${D_\nu }$  becomes the ordinary covariant derivative ${\nabla _\nu }$  when applied to the gauge-invariant tensor EMT.
\begin{equation}
\label{theIdent}
{\nabla _\nu }\left( {{T_\mu}^\nu } \right) = \left( {\frac{{\partial L}}{{\partial {D_\nu }{\bm\phi }}}} \right)\left[ {{D_\mu },{D_\nu }} \right]{\bm\phi } + \bm\Lambda {D_\mu }{\bm\phi }
\end{equation}
The divergence of the canonical energy-momentum tensor (EMT), immediately relates to the Euler-Lagrange equation of motion, and a \textit{commutator} of the gauge covariant derivative. This relation is a consequence of 1) the form of the Lagrangian, 2) the chain rule and the product rule. The non-commuting gauge covariant derivative must be defined as applied to the wave function, and it remains to associate this term with a force.

The divergence of the EMT vanishes when the commutator vanishes (no external forces), and the Euler Lagrange equations are satisfied. In the special case of electromagnetics acting on the wave function in flat space, this ``vanishing divergence'' represents conservation of energy and momentum.\cite{LandauVanishDiverg}

Note the that this relation can be applied to any function of components which supports a derivative. The Lagrangian is not required to be a scalar, nor the indices \textit{tensor} indices. The relation is an identity with the definitions and the functional dependency of the Lagrangian. Mathematically, there is no requirement to be related to Physics!

However, Physics is our domain, and ours to discover the mathematics which allows interpreting the term with the commutator as a generalized Lorentz force which is applied externally.

\section{The generalized Lorentz Force}
We now have the definitions available to show that the electromagnetic field tensor acts on the current to subject the system to the Lorentz force.  Substitute the result of the commutation Eq.~(\ref{gcdComm}) into the identity Eq.~(\ref{theIdent}),
\begin{equation}
\label{divEMT}
{D_\nu }\left( {{T_\mu }^\nu } \right) = i\frac{{\partial L}}{{\partial {D_\nu }{\bm\phi }}}{F_{\mu \nu }}{\bm\phi } + \bm\Lambda {D_\mu }{\bm\phi }
\end{equation}
Substitute the current defined in Eq.~(\ref{current}).
\begin{equation}
\label{divEMTfinal}
\boxed{{D_\nu }\left( {{T_\mu }^\nu } \right) = F_{\mu \nu }{j^\nu } + \bm\Lambda {D_\mu }{\bm\phi }}
\end{equation}
This is the final form of the identity between the divergence of the energy momentum tensor, the generalized Lorentz force, and the Euler-Lagrange equation. In this form, it becomes clear that the external field ${F_{\mu \nu }}$ acts on the current ${j^\mu }$, and so exchanging energy with the system as determined by the divergence of the energy momentum tensor.

It should be noted that the Lorentz force relation, Eq.~(\ref{divEMTfinal}) has been derived without assuming any space-time homogeneity, or symmetry of the Lagrangian.

The generalized Lorentz force in Eq.~(\ref{divEMTfinal}) is a generalization of the electromagnetic Lorentz force equation in Eq.~(\ref{emLorentzForce}).

\section{Vanishing divergence may not imply local continuity}
In electromagnetics, the vanishing divergence implies continuity equations for the components of charged current. However, for non-abelian gauge transformations, a vanishing divergence of the current no longer immediately implies continuity. The additional gauge covariance indicated by the parameter index ``$a$'' in the current $J_a^\mu$. The divergence of the current, Eq.~(\ref{divCurrent}), adds a term including the externally applied potential.
\begin{equation}
  {D_\mu }J_a^\mu  = {\nabla _\mu }J_a^\mu  + A_\mu ^bf_{\;\;ab}^cJ_c^\mu  = 0
\end{equation}
where the tensor covariant derivative is represented by ${\nabla _\mu }$.  The additional term involving the gauge potential prevents the vanishing gauge covariant divergence from becoming a continuity equation.

The covariant divergence is brought into a form exposing ordinary partial derivatives by using an identity for the covariant derivative, Eq.~(\ref{divZeroPartial}).
\begin{equation}
  {D_\mu }J_a^\mu  = \frac{1}{{\sqrt g }}{\partial _\mu }\left( {\sqrt g {j^\mu }} \right) + A_\mu ^bf_{\;\;ab}^cJ_c^\mu  = 0
\end{equation}
This form exposes the partial derivatives in the context of curved space, indicating explicitly that the divergence of partial derivatives fails to vanish due to the gauge potential term, hence continuity fails.

Continuity also fails when applied to a second order tensor such as the EMT, because of the additional term involving metric derivatives.\cite{CarmeliP60}
\begin{equation}
  {\nabla _\nu }\left( {{T_\mu }^\nu } \right) = \frac{1}{{\sqrt { - g} }}\frac{{\partial \left( {{T_\mu }^\nu \sqrt { - g} } \right)}}{{\partial {x^\nu }}} - \tfrac{1}{2}{T^{\nu \beta }}\frac{{\partial {g_{\beta \mu }}}}{{\partial {x^\nu }}} = 0
\end{equation}

Vanishing divergence equations fail to represent differential continuity equations in two important cases, energy-momentum in curved space and non-Abelian gauge fields.

\section{Conclusion}
Two ``divergence-type'' relations, one for the charged current vector, and the other for the energy-momentum tensor motivate the Lie algebra machinery required to construct a gauge covariant derivative of a wave function. The gauge covariant derivative is essential to the definition of charged currents, and their vanishing divergence. The gauge covariant derivative is a means for incorporating an external generalized Lorentz force, and, when this force is zero, the vanishing divergence of energy-momentum tensor.

The definition of the charged current arises from the definition of the gauge transformation, whether or not it is a symmetry of the Lagrangian. The charged current is acted upon by the generalized Lorentz force due to the external field, exchanging energy with the system according to a divergence relation for the energy momentum tensor of the system.

Although the principle of least action is not used here, the two divergence relations provide sufficient motivation for the Euler-Lagrange equations as the equation of motion defined by a Lagrangian. It is interesting that this approach allows consideration of ``off-shell'' equations of motion.

The initial assumption of covariance of the wave function is sufficient to discover all transformation properties of the wave function and its gauge covariant derivative. Further, the assumption of covariance guides the construction of the gauge covariant derivative. The types of covariance considered:  covariant, contravariant, invariant, and gauge field transformation. Of these, only the infinitesimal transformation of the gauge field is not a homogeneous linear transformation.

Lie's Second Theorem follows from the principle of covariance which appears as the requirement that both matrix and parameter indices transform homogeneously in such a way as to preserve the infinitesimal gauge invariance of the generators.

The Lagrangian is assumed to not have an explicit appearance of the external gauge field, since it appears only within the gauge covariant derivative.

Continuity implied by vanishing ``divergence-type'' equations is lost when generalizing the partial derivative to the gauge covariant derivative because of an additional term not generally vanishing.

\begin{acknowledgments}

I thank Jacques Rutschmann for providing just the right amount of skepticism and encouragement.

\end{acknowledgments}

\appendix   

\section{Example:  the wave equation in a gauge field}

Consider an example of the Klein-Gordon wave equation for a complex scalar field $\phi$ interacting with an external electromagnetic field which appears only in the covariant derivative. As is shown in many places, the Lagrangian which yields the Klein-Gordon wave equation of motion, Eq.~(\ref{KGequ}) is,
\begin{equation}
  L\left( {\phi ,{D_\mu }\phi {,}{g_{\mu \nu }}} \right) = {D^\lambda }{\phi ^*}{D_\lambda }\phi  - {M^2}{\phi ^*}\phi 
\end{equation}

However, here we examine the Lagrangian for a wave function consisting of a column matrix of scalars, represented as ${\bm\phi }$  which will yield a wave equation similar to the Klein-Gordon. The following is independent of the specific value of the structure constants which may be substituted later. Again, the Lagrangian has a similar appearance to the Lagrangian for the Klein-Gordon equation.
\begin{equation}
\label{genKGLagr}
  L\left( {{\bm\phi },{D_\mu }{\bm\phi ,}{g_{\mu \nu }}} \right) = {D^\lambda }{{\bm\phi }^\dag }{D_\lambda }{\bm\phi } - {M^2}{{\bm\phi }^\dag }{\bm\phi }
\end{equation}
Apply the gauge covariant derivative to the Lagrangian, and use the chain rule, following the same steps as in the derivation for the energy momentum tensor Eq.~(\ref{chainR2}) then followed by its divergence Eq.~(\ref{divEMTfinal}).
\begin{equation}
\begin{gathered}
  {D_\mu }L = {D_\mu }\left( {{D^\lambda }{{\bm\phi }^\dag }} \right){D_\lambda }{\bm\phi } + {D^\lambda }{{\bm\phi }^\dag }{D_\mu }\left( {{D_\lambda }{\bm\phi }} \right) \\ 
   - {M^2}{D_\mu }\left( {{{\bm\phi }^\dag }} \right){\bm\phi } - {M^2}{{\bm\phi }^\dag }{D_\mu }{\bm\phi }
\end{gathered} 
\end{equation}
Add and subtract the second order derivative, but with swapped indices. Use a commutator bracket,
\begin{equation}
  \left[ {{D_\mu },{D_\lambda }} \right]{\bm\phi  = }{D_\mu }{D_\lambda }{\bm\phi  - }{D_\lambda }{D_\mu }{\bm\phi }
\end{equation}
so that,
\begin{equation}
\begin{gathered}
  {D_\mu }L = \left( {\left[ {{D_\mu },{D_\lambda }} \right]{{\bm\phi }^\dag }} \right){D^\lambda }{\bm\phi  + }{D_\lambda }\left( {{D_\mu }{{\bm\phi }^\dag }} \right){D^\lambda }{\bm\phi } \\ 
   + {D^\lambda }{{\bm\phi }^\dag }\left( {\left[ {{D_\mu },{D_\lambda }} \right]{\bm\phi }} \right) + {D^\lambda }{{\bm\phi }^\dag }{D_\lambda }\left( {{D_\mu }{\bm\phi }} \right) \\ 
   - {M^2}{D_\mu }\left( {{{\bm\phi }^\dag }} \right){\bm\phi } - {M^2}{{\bm\phi }^\dag }{D_\mu }{\bm\phi } \\ 
\end{gathered} 
\end{equation}
Apply the product rule, and rearrange to expose the Euler-Lagrange equations of motion which are the coefficient of the derivative of the field ${D_\lambda }{\bm\phi }$.
\begin{equation}
  \begin{gathered}
  {D_\mu }L = {D_\lambda }\left( {{D_\mu }{{\bm\phi }^\dag }{D^\lambda }{\bm\phi }} \right){ - }\left( {{D_\mu }{{\bm\phi }^\dag }} \right)\left( {{D_\lambda }{D^\lambda }{\bm\phi  + }{M^2}{\bm\phi }} \right) \\ 
   + {D_\lambda }\left( {{D^\lambda }{{\bm\phi }^\dag }{D_\mu }{\bm\phi }} \right) - \left( {{D_\lambda }{D^\lambda }{{\bm\phi }^\dag } + {M^2}{{\bm\phi }^\dag }} \right)\left( {{D_\mu }{\bm\phi }} \right) \\ 
   + \left( {\left[ {{D_\mu },{D_\lambda }} \right]{{\bm\phi }^\dag }} \right){D^\lambda }{\bm\phi } + {D^\lambda }{{\bm\phi }^\dag }\left( {\left[ {{D_\mu },{D_\lambda }} \right]{\bm\phi }} \right) \\ 
\end{gathered} 
\end{equation}
Identify the Euler-Lagrange equation, and its complex conjugate transpose. These are the \textit{generalized} Klein-Gordon equation.
\begin{equation}
\label{genKGequs}
  \begin{gathered}
  {D^\lambda }{D_\lambda }{\bm\phi } + {M^2}{\bm\phi  = 0} \hfill \\
  {D_\lambda }{D^\lambda }{{\bm\phi }^\dag } + {M^2}{{\bm\phi }^\dag }{ = 0} \hfill \\ 
\end{gathered} 
\end{equation}
Substitute these equations, and collect terms under the derivative.
\begin{equation}
  \begin{gathered}
  {D_\mu }L = {D_\lambda }\left( {{D_\mu }{{\bm\phi }^\dag }{D^\lambda }{\bm\phi }{{ + }^\dag }{D_\mu }{\bm\phi }{D^\lambda }{\bm\phi }} \right) \\ 
   + \left( {\left[ {{D_\mu },{D_\lambda }} \right]{{\bm\phi }^\dag }} \right){D^\lambda }{\bm\phi } + {D^\lambda }{{\bm\phi }^\dag }\left( {\left[ {{D_\mu },{D_\lambda }} \right]{\bm\phi }} \right)
\end{gathered} 
\end{equation}
Insert the Kronecker delta, move terms to the left hand side. Rename dummy indices.
\begin{equation}
\label{renamDummy}
  \begin{gathered}
  {D_\nu }\left[ {{D_\mu }{{\bm\phi }^\dag }{D^\nu }{\bm\phi } + {D^\nu }{{\bm\phi }^\dag }{D_\mu }{\bm\phi } - \delta _\mu ^\nu L} \right] =  \hfill \\
   - \left( {\left[ {{D_\mu },{D_\nu }} \right]{{\bm\phi }^\dag }} \right){D^\nu }{\bm\phi } - {D^\nu }{{\bm\phi }^\dag }\left( {\left[ {{D_\mu },{D_\nu }} \right]{\bm\phi }} \right) \hfill
\end{gathered}
\end{equation}

The energy-momentum tensor ${T_{S\mu \nu }}$ for the wave function appears inside the curly bracket.
\begin{equation}
  T_{S\mu }^{\;\;\;\nu } = {D_\mu }{{\bm\phi }^\dag }{D^\nu }{\bm\phi } + {D^\nu }{{\bm\phi }^\dag }{D_\mu }{\bm\phi } - \delta _\mu ^\nu L
\end{equation}
Substitute the commutator results, Eq.~(\ref{gcdComm}) into Eq.~(\ref{renamDummy}). 
\begin{equation}
  {D_\nu }T_{S\mu }^{\;\;\;\nu } = iF_{\mu \nu }^a\left( {{{\bm\phi }^\dag }{{\mathbf{t}}_a}{D^\nu }{\bm\phi } - \left( {{D^\nu }{{\bm\phi }^\dag }} \right){{\mathbf{t}}_a}{\bm\phi }} \right)
\end{equation}

The charged current is defined as,
\begin{equation}
\label{genKGcurrent}
  j_a^\nu  = i\left( {{{\bm\phi }^\dag }{{\mathbf{t}}_a}{D^\nu }{\bm\phi } - \left( {{D^\nu }{{\bm\phi }^\dag }} \right){{\mathbf{t}}_a}{\bm\phi }} \right)
\end{equation}
then recognize the Lorentz force on the right hand side of the following,
\begin{equation}
  {D_\nu }T_{S\mu }^{\;\;\;\nu } = F_{\mu \nu }^aj_a^\nu 
\end{equation}
The divergence of the canonical energy-momentum tensor is equal to the generalized Lorentz force, assuming satisfaction of the Euler-Lagrange equation. As expected, application of a generalized Lorentz force to the charged current results in an exchange of energy with the system.

Consider the Abelian case where the generators all commute with each other so that the structure constants are zero. Mutual commutation allows the generators to be simultaneously diagonalized. We will discover in this interesting case, that the model ``splits up'' and becomes a set of independent complex fields, each obeying the Klein-Gordon equation.

The column matrix form of the wave function may be made explicit by indexing the components so that ${{\phi }_{\left( i \right)}}$ represents a single complex value, and $\left( i \right)$ indicates a matrix index, and \textit{not} a parameter index.
\begin{equation}
{\bm\phi }\left( x \right){ = }\left( {\begin{array}{*{20}{c}}
  {{\phi _{\left( 1 \right)}}\left( x \right)} \\ 
   \vdots  \\ 
  {{\phi _{\left( n \right)}}\left( x \right)} 
\end{array}} \right)
\end{equation}

then substituting into the Lagrangian for the scalar field, Eq.~(\ref{genKGLagr}), find that the Lagrangian becomes a sum of individual Lagrangians.
\begin{equation}
  L = \sum\limits_i {{L_{\left( i \right)}}} 
\end{equation}
Examine how this ``splitting'' comes about. The ${{\bm\phi }^\dag }{\bm\phi }$ term is
\begin{equation}
{{\bm\phi }^\dag }{\bm\phi } = \sum\limits_i {\phi _{\left( i \right)}^*{\phi _{\left( i \right)}}} \end{equation}
The gauge covariant derivative is diagonal since each generator is diagonal.
\begin{equation}
{D_{\left( i \right)\mu }}{\phi _{\left( i \right)}} = {\partial _\mu }{\phi _{\left( i \right)}} - iA_\mu ^a{{\mathbf{t}}_{\left( {ii} \right)a}}{\phi _{\left( i \right)}}
\end{equation}
so that the ``kinetic energy'' term becomes,
\begin{equation}
{D^\lambda }{{\bm\phi }^\dag }{D_\lambda }{\bm\phi } = \sum\limits_i {{D_{\left( i \right)\mu }}\phi _{\left( i \right)}^*{D_{\left( i \right)\mu }}{\phi _{\left( i \right)}}} 
\end{equation}
The gauge covariant derivative does not mix components so that each Lagrangian is independent.
\begin{equation}
{L_{\left( i \right)}} = {D_{\left( i \right)\mu }}\phi _{\left( i \right)}^*{D_{\left( i \right)\mu }}{\phi _{\left( i \right)}} - {M^2}\phi _{\left( i \right)}^*{\phi _{\left( i \right)}}
\end{equation}
Each component satisfies the Euler-Lagrange equation.
\begin{equation}
\Lambda \left( {{L_{\left( i \right)}}} \right) = D_{\left( i \right)}^\lambda {D_{\left( i \right)\lambda }}{\phi _{\left( i \right)}} + {M^2}{\phi _{\left( i \right)}}{ = 0}
\end{equation}

Similarly, the quadratic form of the current $j_a^\nu$ in Eq.~(\ref{genKGcurrent}), splits into a sum of components.
\begin{equation}
 j_a^\nu  = \sum\limits_i {j_{a\left( i \right)}^\nu } 
\end{equation}
where
\begin{equation}
j_{a\left( i \right)}^\nu  = i\left( {\phi _{\left( i \right)}^*{{\mathbf{t}}_{\left( {ii} \right)a}}D_{\left( i \right)}^\nu {\phi _{\left( i \right)}} - \left( {D_{\left( i \right)}^\nu \phi _{\left( i \right)}^*} \right){{\mathbf{t}}_{\left( {ii} \right)a}}{\phi _{\left( i \right)}}} \right)
\end{equation}
and again for the canonical energy momentum tensor,
\begin{equation}
  T_\mu ^{\;\;\;\nu } = \sum\limits_i {T_{\left( i \right)\mu }^{\;\;\;\;\;\;\nu }}
\end{equation}
We see that each complex scalar component of the column matrix of the field ${\phi }$ ``feels'' the same external electromagnetic field, and evolves independently of the other components of the field, without interaction between the scalar field components.
\begin{equation}
  {D_\nu }T_{\left( i \right)\mu }^{\;\;\;\;\;\;\nu } = F_{\mu \nu }^aj_{a\left( i \right)}^\nu 
\end{equation}

An interesting task would be to extend the model, including interactions between the field components.

The charged current is identified, but not yet determined to have a vanishing divergence which connects to a symmetry of the Lagrangian.

\section{Example:  the current}

The current is defined by the infinitesimal gauge transformation Eq.~(\ref{theInfGaugeTransf}) and realized in Eq.~(\ref{current}). Apply the infinitesimal gauge transformation to the Lagrangian in Eq.~(\ref{genKGLagr}), and use the chain rule, following the same steps as in the derivation for the current in then followed by the divergence of the current Eq.~(\ref{divCurrent}).
\begin{equation}
  \delta L = \delta \left( {{D^\lambda }{{\bm\phi }^\dag }} \right){D_\lambda }{\bm\phi } + {D^\lambda }{{\bm\phi }^\dag }\delta \left( {{D_\lambda }{\bm\phi }} \right) - {M^2}\delta {{\bm\phi }^\dag }{\bm\phi } - {M^2}{{\bm\phi }^\dag }\delta {\bm\phi }
\end{equation}
Upon applying the infinitesimal gauge transformation and its complex conjugate, it is immediately apparent that the Lagrangian is gauge invariant since the terms cancel.
\begin{equation}
  \left( {\tfrac{\delta }{{\delta {\theta ^a}}}} \right)L = 0
\end{equation}
Current conservation may be derived from this, but only if we are not so quick to cancel terms. Substitute the variations, Eq.~(\ref{gaugeTranofwaveF}), and do not cancel terms.
\begin{equation}
  \begin{gathered}
  \left( {\tfrac{\delta }{{\delta {\theta ^a}}}} \right)L =  - i{D^\lambda }{{\bm\phi }^\dag }{{\mathbf{t}}_a}{D_\lambda }{\bm\phi } + i{D^\lambda }{{\bm\phi }^\dag }{{\mathbf{t}}_a}{D_\lambda }{\bm\phi } \\ 
   + i{M^2}{{\bm\phi }^\dag }{{\mathbf{t}}_a}{\bm\phi  - }i{M^2}{{\bm\phi }^\dag }{{\mathbf{t}}_a}{\bm\phi } \\ 
\end{gathered} 
\end{equation}
It might seem strange working with terms which sum to zero, but that is exactly the step performed for the general Lagrangian. In each of the two terms containing derivative, create a divergence, then subtracting the extra term required by the product rule.
\begin{equation}
  \begin{gathered}
  \left( {\tfrac{\delta }{{\delta {\theta ^a}}}} \right)L =  - i{D^\lambda }\left( {{{\bm\phi }^\dag }{{\mathbf{t}}_a}{D_\lambda }{\bm\phi }} \right) + i{{\bm\phi }^\dag }{{\mathbf{t}}_a}\left( {{D^\lambda }{D_\lambda }{\bm\phi }} \right) \\ 
   + i{D_\lambda }\left( {{D^\lambda }{{\bm\phi }^\dag }{{\mathbf{t}}_a}{\bm\phi }} \right) - i\left( {{D_\lambda }{D^\lambda }{{\bm\phi }^\dag }} \right){{\mathbf{t}}_a}{\bm\phi } \\ 
   + i{M^2}{{\bm\phi }^\dag }{{\mathbf{t}}_a}{\bm\phi  - }i{M^2}{{\bm\phi }^\dag }{{\mathbf{t}}_a}{\bm\phi } \\ 
\end{gathered} 
\end{equation}
Rearrange to expose the Euler-Lagrange expression, Eq.~(\ref{genKGequs}) for this Lagrangian.
\begin{equation}
\begin{gathered}
  \left( {\tfrac{\delta }{{\delta {\theta ^a}}}} \right)L =  - i{D^\lambda }\left( {{{\bm\phi }^\dag }{{\mathbf{t}}_a}{D_\lambda }{\bm\phi }} \right) + i{{\bm\phi }^\dag }{{\mathbf{t}}_a}\left( {{D^\lambda }{D_\lambda }{\bm\phi } + {M^2}{{\mathbf{t}}_a}{\bm\phi }} \right) \\ 
   + i{D_\lambda }\left( {{D^\lambda }{{\bm\phi }^\dag }{{\mathbf{t}}_a}{\bm\phi }} \right) - i\left( {{D_\lambda }{D^\lambda }{{\bm\phi }^\dag } + {M^2}{{\bm\phi }^\dag }} \right){{\mathbf{t}}_a}{\bm\phi } \\ 
\end{gathered} 
\end{equation}
and the remaining terms are the divergence of the gauge invariant current.
\begin{equation}
  \left( {\tfrac{\delta }{{\delta {\theta ^a}}}} \right)L =  - i{D_\lambda }\left( {{{\bm\phi }^\dag }{{\mathbf{t}}_a}{D^\lambda }{\bm\phi } - \left( {{D^\lambda }{{\bm\phi }^\dag }} \right){{\mathbf{t}}_a}{\bm\phi }} \right)
\end{equation}
so that the current, Eq.~(\ref{genKGcurrent}), is conserved, if the infinitesimal gauge transformation represents a symmetry.
\begin{equation}
  \left( {\tfrac{\delta }{{\delta {\theta ^a}}}} \right)L =  - {D_\lambda }j_a^\lambda  = 0
\end{equation}

This calculation derives from a Lagrangian the equation of motion for charged matter, represented by the complex wave function, in an external field, Eq.~(\ref{genKGequs}). The divergence of the canonical energy momentum tensor equals the Lorentz force as it acts on a current which has a vanishing divergence, as it must, in order to be identified as the charged current interacting with the external gauge field tensor.

\section{Example:  Yang-Mills equations of motion}
The Lagrangian for the electromagnetic field
\begin{equation}
  {L_{{\text{em}}}}\left( {{F^{\kappa \eta }}} \right) = {F^{\kappa \eta }}{F_{\kappa \eta }}
\end{equation}

is a quadratic in field strength. The generalization of the Lagrangian for a gauge field tensor may be taken to be
\begin{equation}
\label{gaugeLagr}
  {L_{{\text{gauge}}}}\left( {F_a^{\kappa \eta }} \right) = F_a^{\kappa \eta }F_{\kappa \eta }^a
\end{equation}
where the parameter index is raised and lowered as needed with the Cartan-Killing metric ${g_{ab}}$.
We will discover the equations of motion, and the energy momentum tensor as part of taking the derivative of the Lagrangian analogous to the first step in Eq.~(\ref{chainR2}). The gauge covariant derivative ${D_\mu }$ is used since the $F_{\mu \nu }^a$ field tensor is \textit{not} gauge invariant.
\begin{equation}
\label{derivEMLagr}
  {D_\mu }{L_{{\text{gauge}}}} = 2{D_\mu }\left( {F_{\kappa \eta }^a} \right)F_a^{\kappa \eta }
\end{equation}

Apply the useful relation for antisymmetric tensors\cite{SchwingerP237} where   indicates the tensor dual (see Appendix:  Definition of the Tensor Dual) of ${F^{\kappa \eta }}$, and ${G_{\kappa \eta }}$ is an arbitrary antisymmetric second order tensor.
\begin{equation}
  \left( {{\nabla ^\eta }{F^{\kappa \mu }}} \right){G_{\kappa \eta }} - \tfrac{1}{2}\left( {{\nabla ^\mu }{F^{\kappa \eta }}} \right){G_{\kappa \eta }} = {\tilde G^{\mu \lambda }}\left( {{\nabla _\nu }\tilde F_{\;\;\lambda }^\nu } \right)
\end{equation}
Move indices, and apply to our case where $G = F$
\begin{equation}
  \left( {{\nabla _\mu }{F_{\kappa \eta }}} \right){F^{\kappa \eta }} = 2\left( {{\nabla _\eta }{F_{\kappa \mu }}} \right){F^{\kappa \eta }} - 2{\tilde F_{\mu \lambda }}\left( {{\nabla _\nu }\tilde F_{\;\;\;\;}^{\nu \lambda }} \right)
\end{equation}
then substitute this relation into Eq.~(\ref{derivEMLagr}), realizing that the same relation holds for second order tensors with a parameter index.
\begin{equation}
  {D_\mu }L = 4{\nabla _\eta }\left( {F_{\kappa \mu }^a} \right)F_a^{\kappa \eta } - 4\tilde F_{\mu \lambda }^a{\nabla _\nu }\left( {\tilde F_a^{\nu \lambda }} \right)
\end{equation}
In first term on the right, pull the second field tensor into the derivative, then subtract the additional term arising from the product rule.
\begin{equation}
  {D_\mu }L = 4{D_\eta }\left( {F_{\kappa \mu }^aF_a^{\kappa \eta }} \right) - 4F_{\kappa \mu }^a{D_\eta }\left( {F_a^{\kappa \eta }} \right) - 4\tilde F_{\mu \lambda }^a{D_\nu }\left( {\tilde F_a^{\nu \lambda }} \right)
\end{equation}
Use the Kronecker delta tensor, and the definition of the Lagrangian.
\begin{equation}
  {D_\eta }\left[ {\tfrac{1}{4}\delta _\mu ^\eta L - F_{\kappa \mu }^aF_a^{\kappa \eta }} \right] =  - F_{\kappa \mu }^a{D_\eta }\left( {F_a^{\kappa \eta }} \right) - \tilde F_{\mu \lambda }^a{D_\nu }\left( {\tilde F_a^{\nu \lambda }} \right)
\end{equation}
Substitute the Lagrangian for the following definition of the energy momentum tensor,
\begin{equation}
  T_{\;\;\mu \,\left( {{\text{gauge}}} \right)}^\eta  = \tfrac{1}{4}\delta _\mu ^\eta F_a^{\kappa \lambda }F_{\kappa \lambda }^a - F_{\kappa \mu }^aF_a^{\kappa \eta }
\end{equation}
and Maxwell's (symmetrized) equations,
\begin{equation}
\label{symMaxwell}
  \begin{gathered}
  j_a^\kappa  = {D_\eta }\left( {F_a^{\eta \kappa }} \right) \hfill \\
  \tilde j_a^\lambda  = {D_\nu }\left( {\tilde F_a^{\nu \lambda }} \right) \hfill \\ 
\end{gathered} 
\end{equation}
to get the divergence of the EMT for the gauge tensor field.
\begin{equation}
\label{divEMTgenElecM}
  {\nabla _\nu }T_{\;\;\mu \,\left( {{\text{gauge}}} \right)}^\nu  = F_{\nu \mu }^aj_a^\nu  + \tilde F_{\nu \mu }^a\tilde j_a^\nu 
\end{equation}
In this symmetrized form, ${\tilde j_\lambda }$ is regarded as a ``magnetic'' current of which is set to zero to obtain Maxwell's equations for electromagnetics.

Notice that we can add the two ``divergence of the EMT'' Eqs.~(\ref{divEMTgenElecM}) and \ref{divEMTfinal}. Assume the Euler-Lagrange equation is satisfied for the wave function, Eq.~(\ref{eulerLagrange})
\begin{equation}
  {\nabla _\nu }\left( {T_{\;\;\mu \,\left( {{\text{gauge}}} \right)}^\nu  + {T_\mu }^\nu } \right) = F_{\nu \mu }^aj_a^\nu  + \tilde F_{\nu \mu }^a\tilde j_a^\nu  + F_{\mu \nu }^aJ_a^\nu 
\end{equation}
Define the sum of the EMTs.
\begin{equation}
  T_{\;\;\mu \,\left( {{\text{total}}} \right)}^\nu  = T_{\;\;\mu \,\left( {{\text{gauge}}} \right)}^\nu  + {T_\mu }^\nu 
\end{equation}
It is tempting to make the following identifications,
\begin{equation}
  \begin{gathered}
  \tilde j_a^\nu  = 0 \hfill \\
  j_a^\nu  =  - J_a^\nu  \hfill \\ 
\end{gathered} 
\end{equation}
so that the sum of the EMTs has a vanishing divergence.
\begin{equation}
  {\nabla _\nu }T_{\;\;\mu \,\left( {{\text{total}}} \right)}^\nu  = 0
\end{equation}
We have constructed an isolated system with these identifications. The gauge field is no longer external with the introduction of a Lagrangian Eq.~(\ref{gaugeLagr}) for that field. In this model, the wave function creates a charged current which is subject to Lorentz forces applied by the gauge field, and at the same time the charged current is the source of the gauge field via Eq.~(\ref{symMaxwell}).

\section{Definition of the Tensor Dual}
Here we pause and define the ``tensor dual'' which is required for the Lorentz force relation. The definition of the ``ordinary'' tensor dual\cite{CarmeliP38} is,
\begin{equation}
  \begin{gathered}
  {{\tilde F}^{\mu \nu }} = \tfrac{1}{2}\tfrac{1}{{\sqrt { - g} }}{\varepsilon ^{\mu \nu \rho \sigma }}{F_{\rho \sigma }} \hfill \\
  {{\tilde F}_{\rho \sigma }} =  - \tfrac{1}{2}\sqrt { - g} {\varepsilon _{\rho \sigma \alpha \beta }}{F^{\alpha \beta }} \hfill \\ 
\end{gathered} 
\end{equation}
where antisymmetric permutation tensors are defined\cite{LovelockP113},
\begin{equation}
  {\varepsilon ^{0123}} = 1
\end{equation}
Application of the dual twice, results in the negative of the original antisymmetric tensor for the space-time metric which has a negative determinant.
\begin{equation}
  - {F_{\mu \nu }} = {\tilde \tilde F_{\mu \nu }}
\end{equation}

\end{document}